\documentclass[11pt,letterpaper]{article}
\usepackage{epsfig,rotating,setspace,latexsym,amsmath,epsf,amssymb,bm}
\usepackage{cite,graphicx,authblk,color,subfigure}

\title{On the Synergistic Benefits of Alternating CSIT for the MISO BC\thanks{E-mail: tandonr@vt.edu, syed@uci.edu, sshlomo@ee.technion.ac.il, poor@princeton.edu. The work of H. V. Poor was supported in part by the Air Force Office of Scientific Research under MURI Grant FA 9550-09-1-0643.}}
\vspace{3cm}
\author[1]{Ravi Tandon}
\author[2]{Syed Ali Jafar}
\author[3]{Shlomo Shamai}
\author[4]{H. Vincent Poor}
\affil[1]{\small Department of ECE, Virginia Tech, Blacksburg, VA, USA.}
\affil[2]{\small Department of EECS, University of California, Irvine, CA, USA.}
\affil[3]{\small Department of EE, Technion, Israel Institute of Technology, Haifa, Israel.}
\affil[4]{\small Department of EE, Princeton University, Princeton, NJ, USA.}

\newtheorem{Theo}{Theorem}
\newtheorem{remark}{Remark}

\begin{document}
\maketitle
\thispagestyle{empty}
\vspace{-0.5cm}
\begin{abstract}
The degrees of freedom (DoF) of the two-user multiple-input single-output (MISO) broadcast channel (BC) are studied under the assumption that the form, $I_i, i=1,2,$ of the channel state information at the transmitter (CSIT) for each user's channel can be either perfect ($P$), delayed ($D$) or not available ($N$), i.e., $I_1, I_2\in\{P,N,D\}$, and therefore the overall CSIT can alternate between the $9$ resulting states $I_1I_2$. The fraction of time associated with CSIT state $I_1I_2$ is denoted by the parameter $\lambda_{I_1I_2}$ and it is assumed throughout that $\lambda_{I_1I_2}=\lambda_{I_2I_1}$, i.e., $\lambda_{PN}=\lambda_{NP}, \lambda_{PD}=\lambda_{DP}, \lambda_{DN}=\lambda_{ND}$.  Under this  assumption of symmetry, the main contribution of this paper is a complete characterization of the DoF region of the two user MISO BC with alternating CSIT. Surprisingly, the DoF region is found to depend 
only on the marginal probabilities $(\lambda_P, \lambda_D,\lambda_N)=\left(\sum_{I_2}\lambda_{PI_2},\sum_{I_2}\lambda_{DI_2}, \sum_{I_2}\lambda_{NI_2}\right)$, $I_2\in\{P,D,N\}$, which represent the fraction of time that any given user (e.g., user 1) is associated with perfect, delayed, or no CSIT, respectively. As a consequence, the DoF region with all 9 CSIT states, $\mathcal{D}(\lambda_{I_1I_2}:I_1,I_2\in\{P,D,N\})$, is the same as the DoF region with only 3 CSIT states $\mathcal{D}(\lambda_{PP}, \lambda_{DD}, \lambda_{NN})$, under the same marginal distribution of CSIT states, i.e.,  $(\lambda_{PP}, \lambda_{DD},\lambda_{NN})=(\lambda_P,\lambda_D,\lambda_N)$. The sum-DoF value can be expressed as $\mbox{DoF}=\min\left(\frac{4+2\lambda_P}{3}, 1+\lambda_P+\lambda_D\right)$, from which one can uniquely identify the minimum required marginal CSIT fractions to achieve any target DoF value as $(\lambda_P,\lambda_D)_{\min}=\left(\frac{3}{2}\mbox{\tiny DoF}-2,1-\frac{1}{2}\mbox{\tiny DoF}\right)$ when $\mbox{DoF}\in\big[\frac{4}{3},2\big]$ and $(\lambda_P,\lambda_D)_{\min}=(0,(\mbox{\tiny DoF}-1)^+)$ when $\mbox{DoF}\in\big[0,\frac{4}{3}\big)$. The results highlight the synergistic benefits of alternating CSIT and the tradeoffs between various forms of CSIT for any given DoF value. 
\end{abstract}
\newpage
\section{Introduction}
The availability of channel state information at transmitters (CSIT) is a key ingredient for  interference management techniques \cite{Jafar:Tutorial}. It affects not only  the capacity but also the degrees of freedom  (DoF) of wireless networks. Perhaps the simplest setting that exemplifies the critical role of CSIT is the two-user vector broadcast channel, also known as the multiple input single output broadcast channel (MISO BC), in which a transmitter equipped with two antennas sends independent messages to two receivers, each equipped with a single antenna. Degrees of freedom characterizations for the MISO BC are available under a variety of CSIT models, including full (perfect and instantaneous) CSIT \cite{MIMOBC}, no CSIT \cite{CaireShamai,NoCSITJafar,VV:NOCSIT,JafarGoldsmithIsotropic},  delayed CSIT \cite{MaddahAli-Tse:DCSI-BC,VV:DCSI-BC}, compound CSIT \cite{Weingarten_Shamai_Kramer,CompoundJafar,MAMA:CompoundBC}, quantized CSIT \cite{Jindal_BCFB, Caire_Jindal_Shamai,Kobayashi_Caire_Jindal},  mixed (perfect delayed and partial instantaneous) CSIT \cite{JafarTCBC,KobayashiTCBC,EliaMixed}, asymmetric CSIT (perfect CSIT for one user, delayed CSIT for the other) \cite{Jafar_corr, RetroIA} and with knowledge of only the channel coherence patterns available to the transmitter \cite{Jafar_corr, Wang_Gou_Jafar}. Yet, the understanding of the role of CSIT for the MISO BC is far from complete, even from a DoF perspective, as exemplified by the Lapidoth-Shamai-Wigger conjecture \cite{LSWConjecture}, which is but one of the many open problems along this research avenue. 

In this work we focus on an aspect of CSIT that has so far received little direct attention -- \emph{that it can vary over time}. Consider the MISO BC for the case in which perfect CSIT is available for one user and no CSIT is available for the other user. Incidentally, the DoF are unknown for this problem. Now, staying within the assumption of full CSIT for one user and none for the other,  suppose we allow the CSIT to vary, in the sense that half the time we have full CSIT for user 1 and none for user 2, and for the remaining half of the time we have full CSIT for user 2 and none for user 1. This is one example of what we call the \emph{alternating CSIT} setting. In general terms, the defining feature of the alternating CSIT problem is a joint consideration of multiple CSIT states.

We motivate the  alternating CSIT setting by addressing three natural questions --- 1) is it practical, 2) is it a trivial extension, and 3) is it desirable/beneficial, relative to the more commonly studied non-alternating/fixed CSIT settings?

To answer the first question, we note that alternating CSIT may be already practically unavoidable due to the time varying nature of wireless networks. However, more interestingly, the form of CSIT may also be \emph{deliberately} varied as a design choice, often with little or no additional  overhead. For example, acquiring perfect CSIT for one user and none for the other for half the time and then switching the role of users for the remaining half of the time, carries little or no additional overhead relative to the non-alternating case in which perfect CSIT is acquired for the same user for the entire time while no CSIT is obtained for the other user. Thus, alternating CSIT is as practical as the non-alternating CSIT setting. 

The second question relates to the novelty of the alternating CSIT setting with respect to the non-alternating CSIT setting. Is the former just a direct extension of the latter? As we will show in this work, this is not the case. Surprisingly, we find that the lack of a direct relationship between the alternating and non alternating settings  works in our favor. Indeed, we are able to solve the alternating CSIT DoF problem in several cases for which the non-alternating case remains open.  In particular, this includes the above mentioned case of full CSIT for one user and none for the other. As mentioned previously, for this problem the DoF remain open in the  non-alternating CSIT setting. However, we are able to find the DoF for the same problem under the alternating CSIT assumption.

The third question, whether there is a benefit of alternating CSIT relative to non-alternating CSIT, is perhaps   the most interesting question. Here, we will show that the constituent fixed-CSIT settings in the alternating CSIT problem are inseparable (for more on separability, see \cite{Cadambe_Jafar_inseparable, Lalitha_inseparable,Cadambe_Jafar_MACZBC}), so that the DoF of the alternating CSIT setting can be strictly larger than a proportionally weighted combination of the DoF values of the constituent fixed-CSIT settings. We call this the \emph{synergistic DoF gain} of alternating CSIT. As we will show in this work, the benefits of alternating CSIT over non-alternating CSIT can be quite substantial.

{\it Related work:} In terms of the constituent fixed-CSIT schemes, this work is related to most prior studies of the MISO BC DoF. While several recent works on mixed CSIT models, such as \cite{JafarTCBC,KobayashiTCBC,EliaMixed}, also jointly consider multiple forms of CSIT, it is noteworthy that these works are fundamentally distinct as in \cite{JafarTCBC,KobayashiTCBC,EliaMixed},  the multiple forms of CSIT are assumed to be simultaneously present in what ultimately amounts to a fixed-CSIT setting, as opposed to the alternating  CSIT setting considered in this work. More closely related to our setting, are the recent works in \cite{Xu_Andrews_Jafar} and \cite{NamyoonHeath} which involve alternating perfect and delayed CSIT models.  In particular, the three receiver MISO BC with two transmit antennas is studied in \cite{NamyoonHeath}, leading to an interesting observation that the presence of a third user, even with only two transmit antennas, can strictly increase the DoF. 

{\it Organization:} Our model of MISO broadcast channel with alternating CSIT is described in Section \ref{Sec:Model}. In Section \ref{Sec:Results}, we present the DoF region of the MISO BC under alternating CSIT and highlight several aspects and interpretations of the results. In Section \ref{Sec:CS}, we present constituent encoding schemes which highlight the benefits of alternating CSIT.
Achievability of the DoF region with alternating CSIT is presented in Section \ref{Sec:Achievability} and the converse is presented in Section \ref{Sec:Converse}. 

\section{System Model}\label{Sec:Model}
A two user MISO BC is considered, in which a transmitter (denoted as $Tx$) equipped with
two transmit antennas wishes to send independent messages $W_{1}$ and  $W_2$, to two receivers (denoted as $Rx_1$, and $Rx_2$, respectively), and each receiver is equipped with a single antenna. The input-output relationship is given as
\begin{align}
Y(t)&= H(t)X(t) + N_{y}(t)\\
Z(t)&= G(t)X(t) + N_{z}(t),
\end{align}
where $Y(t)$ (resp. $Z(t)$) is the channel output at $Rx_1$ (resp. $Rx_2$) at time $t$, $X(t)=[x_{1}(t) \quad x_{2}(t)]^{T}$ is the $2\times 1$ channel input
which satisfies the power constraint $E[||X(t)||^{2}]\leq P$, and $N_{y}(t), N_{z}(t)\sim \mathcal{CN}(0,1)$ are circularly symmetric complex additive white Gaussian noises at receivers $1$ and $2$ respectively. The $2\times 1$ channel vectors $H(t)$ (to receiver $1$) and $G(t)$ (to receiver $2$)
are independent and identically distributed (i.i.d.) with continuous distributions, and are also i.i.d. over time.
The rate pair $(R_{1},R_{2})$, with $R_{i}= \log(|W_{i}|)/n$, where $n$ is the number of channel uses, is achievable if the probability of decoding error for $i=1,2$
can be made arbitrarily small for sufficiently large $n$. We are interested in the degrees of freedom region $\mathcal{D}$, defined as the set of all achievable pairs $(d_{1},d_{2})$ with $d_{i}=\lim_{P\rightarrow \infty} \frac{R_{i}}{\log(P)}$.

While a  variety of CSIT models are conceivable, here we identify the two most important characteristics of CSIT as  --- 1) precision, and 2) delay. Based on these two characteristics we identify three  forms of CSIT to be considered in this work.
\begin{enumerate}
\item {\it Perfect CSIT ($P$)}: Perfect CSIT, or $P$, denotes those instances in which CSIT is available  instantaneously and with infinite precision.
\item {\it Delayed CSIT ($D$)}: Delayed CSIT, or $D$, denotes those instances in which CSIT is available with infinite precision but only after such delay that it is independent of the current channel state.
\item {\it No CSIT ($N$)}: No CSIT, or $N$, denotes those instances in which no CSIT is available. The users' channels are  statistically indistinguishable in this case.
\end{enumerate}
The CSIT state of user 1, $I_1$, and the CSIT state of user 2, $I_2$, can each belong to any of these three cases,
$$I_1, I_2\in\{P,D,N\},$$ 
giving us a total of 9 CSIT states $I_1I_2\in\{PP, PD, DP, PN,NP,DD, DN,ND,NN\}$ for the two user MISO BC.  Further, let us denote by $\lambda_{I_1I_2}$ the fraction of time that the state $I_1I_2$ occurs, so that 
\begin{eqnarray}
\lambda_{PP}+\lambda_{PD}+\lambda_{DP}+\lambda_{PN}+\lambda_{NP}+\lambda_{DD}+\lambda_{DN}+\lambda_{ND}+\lambda_{NN}=1\label{eq:lambdasum}.
\end{eqnarray}
We will assume throughout this paper that $\lambda_{I_1I_2}=\lambda_{I_2I_1}$. Specifically, 
\begin{eqnarray}
\lambda_{PD}&=&\lambda_{DP}\\
\lambda_{PN}&=&\lambda_{NP}\\
\lambda_{DN}&=&\lambda_{ND}\label{eq:lambdasym}.
\end{eqnarray}
This assumption is justified by the inherent symmetry of the problem, e.g., it is easy to see that if DoF were to be optimized subject to a symmetric CSIT cost constraint (the cost for acquiring CSIT state $I_1I_2$ equals the cost of $I_2I_1$) then the optimal choice of CSIT states will always satisfy the property $\lambda_{I_1I_2}=\lambda_{I_2I_1}$. 
Furthermore, we assume that both the receivers have perfect \emph{global} channel state information. 

\bigskip

\noindent{\bf Problem Statement:} Given the probability mass function (pmf), ${\bf\lambda}_{I_1I_2}$, 
 the problem is to characterize the degrees-of-freedom  region $\mathcal{D}({\bf{\lambda}}_{I_1I_2})$.

\bigskip

\section{Main Results and Insights}\label{Sec:Results}
Starting with the $9$ parameters $\lambda_{I_1I_2}$, even if we use the 4 constraints (\ref{eq:lambdasum})-(\ref{eq:lambdasym}) to eliminate $4$ parameters (say, $\lambda_{DP}, \lambda_{NP}, \lambda_{ND}, \lambda_{NN}$), we are still left with $5$ free parameters ($\lambda_{PP}, \lambda_{PD}, \lambda_{DD}, \lambda_{PN},\lambda_{DN})$, and a  challenging task of characterizing the DoF region which is a function of  these $5$ remaining parameters, i.e., a mapping from a region in $\mathbb{R}^5$ to a region in $\mathbb{R}^2$. While such a problem can easily become intractable or at least extremely cumbersome,  it turns out --- rather serendipitously ---  to be not only  completely solvable but also surprisingly easy to describe. 

\subsection{Main Result}
We start with the main result, stated in the following theorem.

\begin{Theo}\label{Theorem1}
The DoF region $\mathcal{D}(\lambda_{I_1I_2})$, for the two user MISO BC with alternating CSIT  is given by the set of non-negative pairs $(d_{1},d_{2})$ that satisfy
\begin{align}
d_{1}&\leq 1\\
d_{2}&\leq 1\\
d_{1}+2d_{2}&\leq 2+\lambda_{PP}+ \lambda_{PD}+ \lambda_{PN}\\
2d_{1}+d_{2}&\leq 2+\lambda_{PP}+\lambda_{PD}+ \lambda_{PN}\\
d_{1}+d_{2}&\leq 1+ \lambda_{PP}+ 2\lambda_{PD}+\lambda_{DD}+ \lambda_{PN}+\lambda_{DN}.
\end{align}
\end{Theo}
The achievability proof for Theorem \ref{Theorem1} is presented in Section \ref{Sec:Achievability}, and the converse proof is detailed in Section \ref{Sec:Converse}.

Note the dependence of the DoF region in Theorem 1 on the 5 remaining parameters $\lambda_{PP}$, $\lambda_{PD}$, $\lambda_{DD}$, $\lambda_{PN}$, $\lambda_{DN}$. As remarkable as the simplicity of the DoF region description in Theorem \ref{Theorem1} may be, it is possible to simplify it even further, in terms of only two \emph{marginal} parameters -- $\lambda_P$ and $\lambda_D$. This simplification and associated insights are presented next through a set of remarks.
\begin{remark}{\bf [Representation in terms of Marginals]}\label{Rem1}
The DoF region in Theorem \ref{Theorem1} can also be expressed as follows:
\begin{align}
d_{1}&\leq 1\label{Alt1}\\
d_{2}&\leq 1\label{Alt2}\\
d_{1}+2d_{2}&\leq 2+\lambda_{P}\label{Alt3}\\
2d_{1}+d_{2}&\leq 2+\lambda_{P}\label{Alt4}\\
d_{1}+d_{2}&\leq 1+ \lambda_{P}+ \lambda_{D},\label{Alt5}\
\end{align}
where $\lambda_{P}$ and $\lambda_{D}$ defined below denote the {\bf{total}} fraction of time that perfect and delayed CSIT, respectively, are associated with a  user:
\begin{align}
\lambda_{P}&\triangleq \lambda_{PP} + \lambda_{PD} + \lambda_{PN}\\
\lambda_{D}&\triangleq \lambda_{DD} + \lambda_{PD} + \lambda_{DN}.
\end{align}
Note that these two marginal fractions satisfy
\begin{align}
\lambda_{P}+\lambda_{D}+\lambda_{N}=1,
\end{align}
where $\lambda_{N}=\lambda_{NN} + \lambda_{PN} + \lambda_{DN}$ is the total fraction of time that no CSIT is associated with a user. 
\end{remark}

\begin{figure}[t]
  \centering
\includegraphics[width=12.2cm]{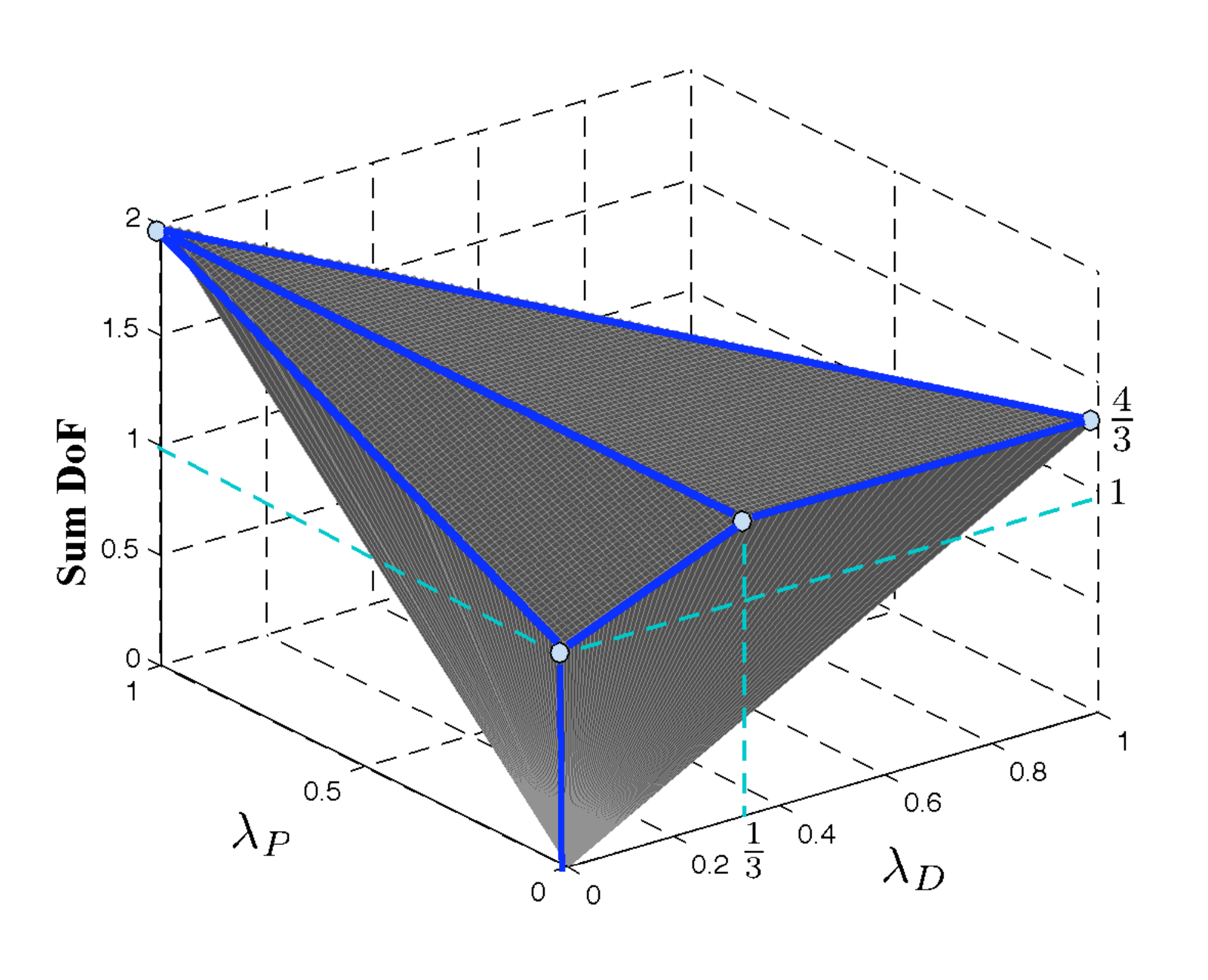}
\caption{Sum DoF as a function of $(\lambda_{D},\lambda_{P})$.}\label{Fig:sumDoF3d}
\end{figure}

\begin{remark}\label{Rem2}{\bf [Same-Marginals Property]}
From Remark \ref{Rem1}, we make a surprising observation. Given any alternating CSIT setting considered in this work, i.e., given any $\lambda_{I_1,I_2}$, there exists an {\bf equivalent}  alternating CSIT problem,  having only three states: PP, DD and NN, with fractions $\lambda_{P}, \lambda_{D}$, and  $\lambda_{N}$ as defined above. The two are equivalent in the sense that they have the same DoF regions. Thus, all alternating CSIT settings considered in this work can be reduced to only symmetric CSIT states with the same marginals, without any change in the DoF region. The sum DoF as a function of $(\lambda_{D},\lambda_{P})$, where $\lambda_{N}=1-\lambda_{P}-\lambda_{D}$ is shown in Figure \ref{Fig:sumDoF3d}. 
\end{remark}
This equivalence, which greatly simplifies the representation of the DoF region, remains rather mysterious because we have not found an argument that could establish this equivalence \emph{a priori}. The equivalence is only evident after Theorem \ref{Theorem1} is obtained, which allows us to simplify the statement of the theorem, but does not simplify the proof of the theorem. Nevertheless, the possibility of a general relationship along these lines is intriguing.

\begin{remark}{\bf [Sum-DoF]}
From (\ref{Alt1})-(\ref{Alt5}), we can write the sum DoF as follows:
\begin{align}
d_{1}+d_{2}&= \min \left(\frac{4+2\lambda_{P}}{3}, 1+\lambda_{P}+\lambda_{D}\right)\label{eq:sumDoF1}\\
&= 2-\frac{2\lambda_{N}}{3} - \frac{\max(\lambda_{N}, 2\lambda_{D})}{3},\label{eq:sumDoF2}
\end{align}
where we used the fact that $\lambda_{P}+\lambda_{D}+\lambda_{N}=1$.
\end{remark}

\begin{remark}{\bf [Cost of Delay]}
It is interesting to contrast the two different forms of CSIT, delayed versus perfect.
From (\ref{eq:sumDoF1}) and (\ref{eq:sumDoF2}) we notice that, depending on the following condition:
\begin{eqnarray}
\lambda_D\geq\frac{\lambda_N}{2}\label{eq:condition},
\end{eqnarray}
we have two very distinct observations. We note that in the region where (\ref{eq:condition}) is true,  delayed CSIT is interchangeable with no CSIT, because the DoF depends only on $\lambda_P$. Here,  delay makes CSIT useless. On the other hand, in the region where $\lambda_D<\frac{\lambda_N}{2}$,  delayed CSIT is as good as perfect CSIT. 
\end{remark}
\begin{remark}{\bf [Minimum Required CSIT for a DoF value]}
This tradeoff between marginal $\lambda_P$ and $\lambda_D$ is explicitly illustrated in Fig. \ref{Fig:CSITrequired}. The most efficient point, in terms of marginal CSIT required to achieve any given value of DoF, is uniquely identified to be the bottom  corner of the left most edge (highlighted corner in Fig. \ref{Fig:CSITrequired}) of the corresponding trapezoid. Note that any other feasible CSIT point involves either redundant CSIT or unnecessary ``instantaneous" CSIT requirements when delayed CSIT would have sufficed just as well. For example, following are the minimum CSIT requirements for various sum-DoF target values:
\allowdisplaybreaks
\begin{eqnarray*}
\mbox{DoF}=\frac{4}{3}&\Rightarrow&(\lambda_P,\lambda_D)=\left(0,\frac{1}{3}\right)\\
\mbox{DoF}=\frac{3}{2}&\Rightarrow&(\lambda_P,\lambda_D)=\left(\frac{1}{4},\frac{1}{4}\right)\\
\mbox{DoF}=\frac{8}{5}&\Rightarrow&(\lambda_P,\lambda_D)=\left(\frac{2}{5},\frac{1}{5}\right)\\
\mbox{DoF}=\frac{5}{3}&\Rightarrow&(\lambda_P,\lambda_D)=\left(\frac{1}{2},\frac{1}{6}\right)\\
\mbox{DoF}=2&\Rightarrow&(\lambda_P,\lambda_D)=\left(1,0\right).
\end{eqnarray*}
In fact, a general expression for the minimum CSIT required to achieve a sum-DoF value is easily evaluated to be
\begin{eqnarray}
\left(\lambda_P,\lambda_D\right)_{\min}&=&\left\{
\begin{array}{cc}
\left(\frac{3}{2}\mbox{\small DoF}-2,1-\frac{1}{2}\mbox{\small DoF}\right),&\mbox{ \small DoF} \in[\frac{4}{3},2]\\
(0,({\small DoF}-1)^+)&\mbox{ \small DoF}\in[0,\frac{4}{3}).
\end{array}
\right.
\end{eqnarray}

\begin{figure}[t]
  \centering
\includegraphics[width=12.2cm]{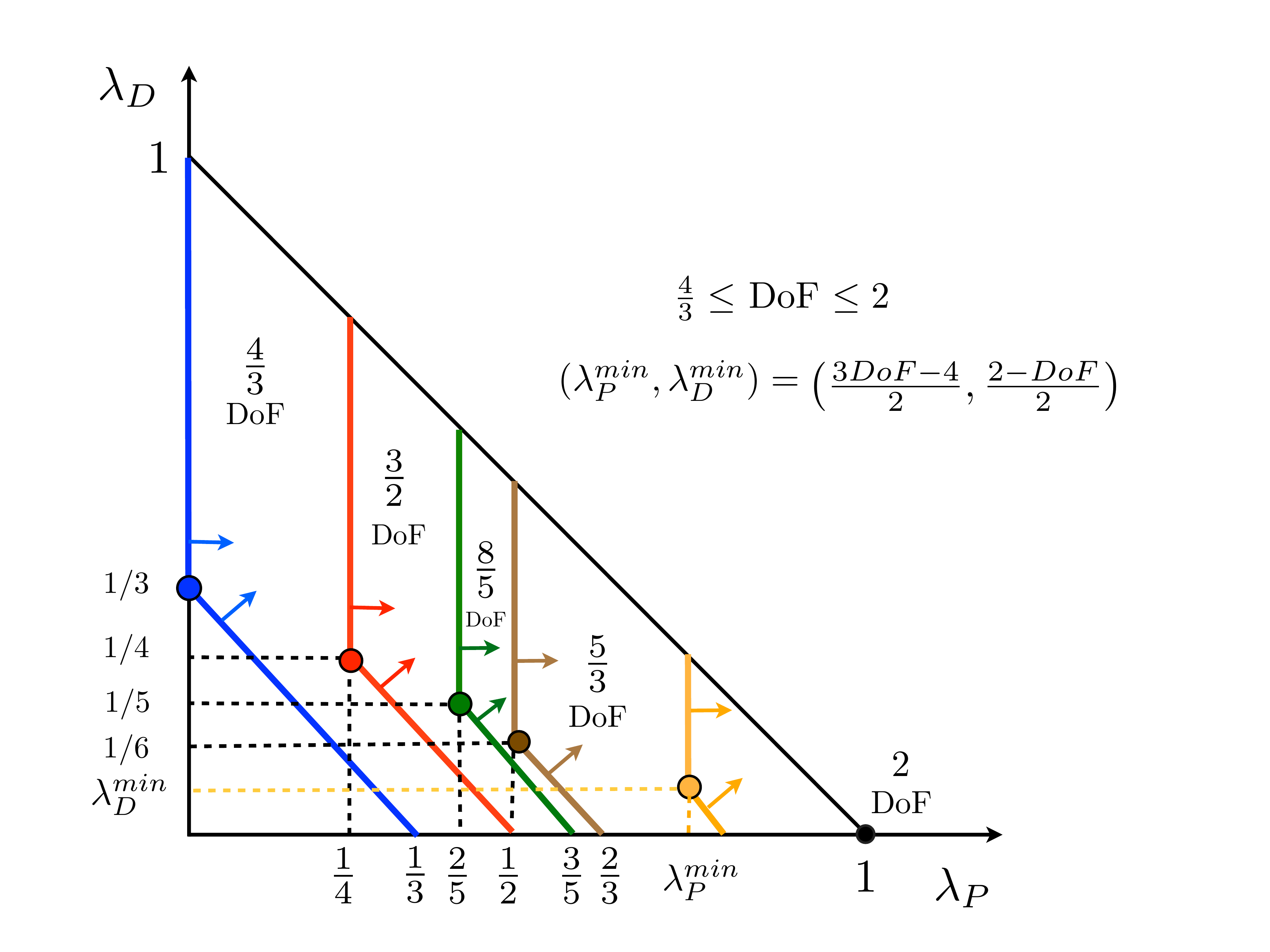}
\caption{Tradeoff between Delayed and Perfect CSIT.}\label{Fig:CSITrequired}
\end{figure}

\end{remark}
\subsection{Synergistic Benefits}
As mentioned previously, the most interesting aspects of the alternating CSIT problem are the synergistic DoF gains. Representative examples of this phenomenon are presented next.
\begin{itemize}
\item {\bf Example 1:} Consider the non-alternating CSIT setting, $PD$, in which perfect CSIT is available for one user and delayed CSIT is available for the other user. It has been shown in \cite{RetroIA} that this setting has $3/2$ DoF. Now, let us make this an alternating CSIT setting. Suppose that half of the time the CSIT is of the form $PD$ and remaining half of the time, the CSIT is of the form $DP$. From the main result stated in Theorem \ref{Theorem1}, it is easy to see that the optimal DoF value is now increased to $5/3$. This is an example of a synergistic DoF gain from alternating CSIT. Figure \ref{comparisonfigure} shows the DoF regions corresponding to the three fixed-CSIT states -- $DD, PD$ and $DP$; and the  DoF region resulting by permitting alternation between states $PD$ and $DP$ in which each state occurs for half of the total communication period. This result also highlights the inseparability of operating over such CSIT states and shows that by jointly coding across these states, thereby collaboratively using the CSIT distributed over time, significant gains in DoF can be achieved. 
\begin{figure}[h]
  \centering
\includegraphics[width=10.4cm]{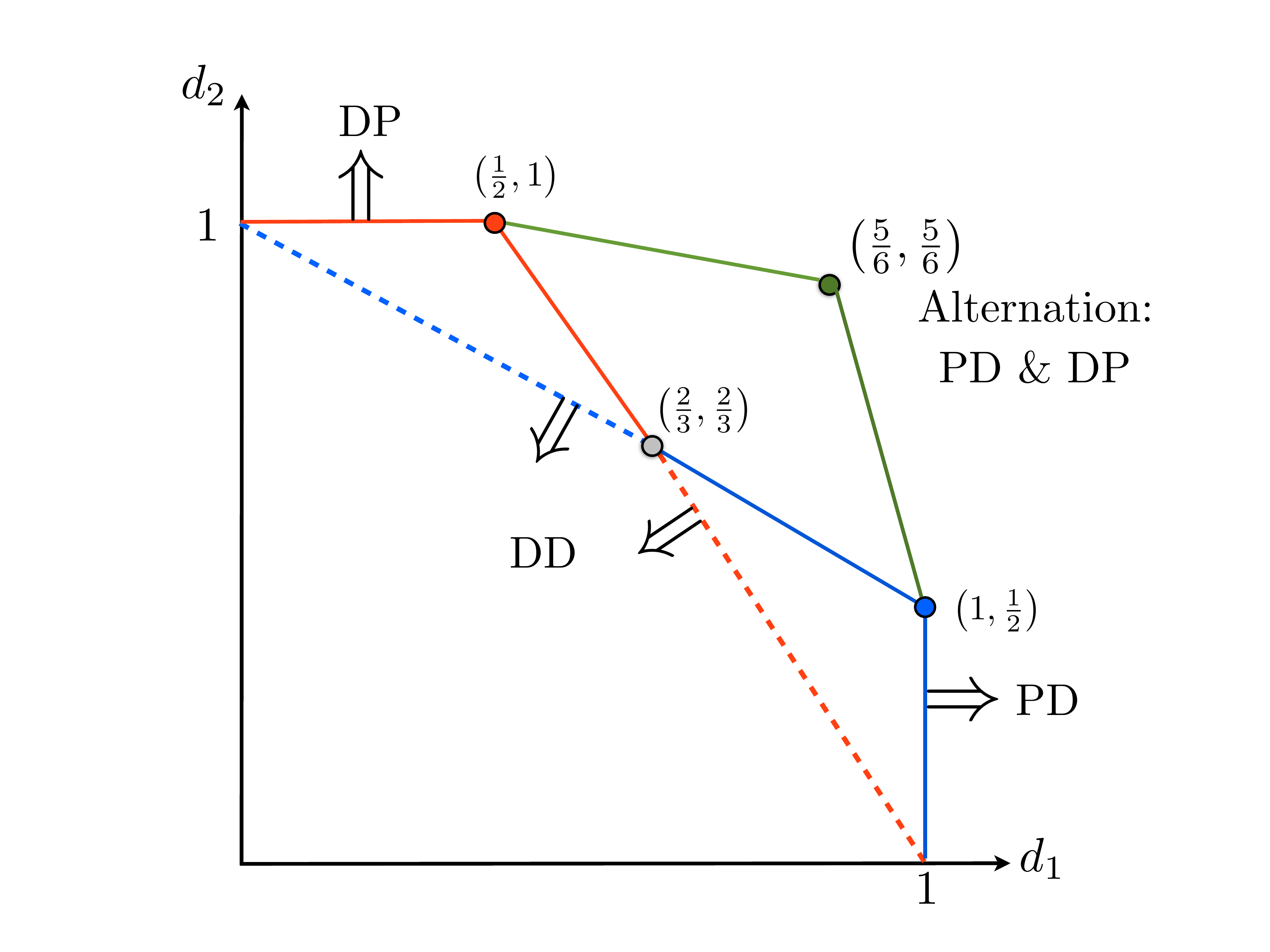}
\vspace{-0.4cm}
\caption{DoF gain via alternating CSIT.}\label{comparisonfigure}
\vspace{-0.4cm}
\end{figure}

\item {\bf Example 2:}  Another interesting example for which alternating CSIT provides provable DoF gains over non-alternating CSIT is the case when states $DD, PN$ and $NP$ are present. Individually, the optimal DoF for $DD$ state is $4/3$ as shown in 
\cite{MaddahAli-Tse:DCSI-BC}. For the $PN$ and $NP$ states, the optimal DoF value is not known; however an upper bound of $3/2$ can be readily established.  In contrast, if alternation is permitted among $DD, PN$ and $NP$, according to $(\lambda_{DD}, \lambda_{PN}, \lambda_{NP})=(\frac{1}{5}, \frac{2}{5},\frac{2}{5})$, then the optimal DoF value is $8/5$, which is larger than both $4/3$ and $3/2$, thereby showing strict synergistic gains made possible by alternating CSIT. 

\item {\bf Example 3:} As  mentioned above, the DoF value is not known individually for fixed-CSIT state $PN$. In fact, it is our conjecture that for fixed-CSIT state $PN$, the optimal DoF value is only $1$. However,  in the alternating CSIT setting, if the states $PN$ and $NP$ are present for equal fractions of the time, then $3/2$  is the optimal DoF value. 

\item{\bf Example 4:} Interestingly enough, the Maddah-Ali and Tse (henceforth referred as MAT) scheme \cite{MaddahAli-Tse:DCSI-BC}, or rather the alternative version of it presented in \cite{KobayashiTCBC}, may also be seen as an alternating CSIT scheme that achieves $\frac{4}{3}$ DoF with $(\lambda_{DD}, \lambda_{NN})=(\frac{1}{3},\frac{2}{3})$. Since the DoF of the $DD$ setting by itself is $\frac{4}{3}$ and the DoF of the $NN$ setting is $1$,  and $\frac{4}{3}>\frac{1}{3}\left(\frac{4}{3}\right)+\frac{2}{3}(1)$, the synergistic gains are evident here as well.

\end{itemize}

We conclude this section by highlighting some of key aspects of the achievability and converse proofs. The converse proofs are inspired by the techniques developed for mixed CSIT configurations in \cite{JafarTCBC} but also include some novel elements. A simple setting that highlights the novel aspects of the converse proof may be the case in which $(\lambda_{PN}, \lambda_{NP})=(1/2, 1/2)$.
For the achievability proof, the main challenge lies in identifying the core constituent schemes. In particular, core constituent schemes achieving DoF values of $3/2, 5/3$ and $8/5$ by using {\emph{minimal}} CSIT under various CSIT states are fundamental to the achievability of the DoF region. These constituent schemes are the topic of the next section.

\newpage

\section{Constituent Schemes}\label{Sec:CS}
In proving the achievability of the respective DoF regions, we first present so called constituent encoding 
schemes that form the key building blocks for the achievability of the region stated in Theorem \ref{Theorem1}.
Furthermore, through these constituent encoding schemes, the benefits of alternating CSIT states can be easily appreciated. 

\subsection{Scheme achieving $1$ DoF}
Achieving $1$ DoF requires no CSIT; and thus any state can be used for this purpose.
We denote the scheme achieving $1$ DoF as follows: 
\begin{itemize}
\item $S^{1}$: uses the state NN and achieves $(d_{1},d_{2})=(1,0)$.
\end{itemize}

\subsection{Scheme achieving $2$ DoF}
The only scheme that achieves $2$ DoF corresponds to the state PP, i.e., when the transmitter has
 perfect CSIT from both receivers. This is achievable via zero-forcing. We denote this scheme as follows:
\begin{itemize}
\item $S^{2}$: uses the state PP and achieves $(d_{1},d_{2})=(1,1)$.
\end{itemize}
 
\subsection{Schemes achieving $4/3$ DoF}
The following schemes achieve $4/3$ DoF:
\begin{itemize}
\item $S^{4/3}_{1}$: using DD and achieving $(d_{1},d_{2})=\left(\frac{2}{3},\frac{2}{3}\right)$. 

This is the scheme presented in \cite{MaddahAli-Tse:DCSI-BC} and achieves sum DoF of $4/3$ as follows: at $t=1$, the transmitter sends two symbols $(u_{1},u_{2})$ intended for receiver $1$; this step delivers a useful information symbol at receiver $1$ and creates side-information at receiver $2$. By a useful information symbol for receiver $1$, we refer to a random linear combination of $u_{1}$ and $u_{2}$.  Similarly, at $t=2$, the transmitter sends two symbols $(v_{1},v_{2})$ intended for receiver $2$; delivering a useful symbol at receiver $2$ while creating side-information at receiver $1$. Due to delayed CSIT, the transmitter can reconstruct the side-information symbols created at $t=1,2$. At $t=3$, the transmitter sends a linear combination of these side-information symbols. After $t=3$, each receiver, upon receiving this linear combination, can remove the interference by using its past overheard information. Therefore, $4/3$ DoF is achievable.

\item $S^{4/3}_{2}$: using DD, NN for fractions $\left(\frac{1}{3}, \frac{2}{3}\right)$ and achieving $(d_{1},d_{2})=\left(\frac{2}{3},\frac{2}{3}\right)$. 

We show this scheme by a modification of the MAT scheme described next.  At $t=1$, the transmitter sends $u_{1}+v_{1}$ on the first antenna and $u_{2}+v_{2}$ on the second antenna. Channel outputs at $t=1$ are as follows: receiver $1$ obtains $A_{1}(u_{1},u_{2})+ B_{1}(v_{1},v_{2})$, whereas receiver $2$ obtains $A_{2}(u_{1},u_{2})+ B_{2}(v_{1},v_{2})$. Via delayed CSIT from $t=1$, the transmitter can reconstruct $B_{1}(v_{1},v_{2})$ and $A_{2}(u_{1},u_{2})$ within noise distortion. At $t=2$, it transmits $A_{2}(u_{1},u_{2})$ to both receivers using one antenna and at $t=3$, it transmits $B_{1}(v_{1},v_{2})$ to both receivers. This scheme also achieves a DoF of $4/3$. The interesting aspect is that delayed CSIT from both receivers is required only at $t=1$; however no CSIT is required from $t=2,3$. Thus, by alternation between (DD, NN) for fractions $(1/3,2/3)$, $4/3$ DoF is achievable. This modification of the original MAT scheme is also mentioned in \cite{KobayashiTCBC} and \cite{MaddahAli-Tse:DCSI-BC}.

\item $S^{4/3}_{3}$: using DN, ND for fractions $\left(\frac{1}{2}, \frac{1}{2}\right)$ and achieving $(d_{1},d_{2})=\left(\frac{2}{3},\frac{2}{3}\right)$.

\item $S^{4/3}_{4}$: using DN, ND, NN for fractions $\left(\frac{1}{3},\frac{1}{3}, \frac{1}{3}\right)$ and achieving $(d_{1},d_{2})=\left(\frac{2}{3},\frac{2}{3}\right)$.

We now present the combined explanation of the schemes $S^{4/3}_{3}$ and $S^{4/3}_{4}$. In the original MAT scheme mentioned for $S^{4/3}_{1}$, after $t=1$, the transmitter requires CSIT only from receiver $2$; after $t=2$, the transmitter  requires CSIT only from receiver $1$ and at $t=3$, the transmitter requires no CSIT. From this observation, we note that the original assumption of global delayed CSIT  can be relaxed to one in which the transmitter can choose to select the available CSIT from a set of three states: state ND--no CSIT from receiver $1$ and delayed CSIT from receiver $2$; state DN--delayed CSIT from receiver $1$ and no CSIT from receiver $2$; and state NN--no CSIT from either of the receivers. If in addition, it is required that these states have to be chosen for an equal fraction  (i.e., one-third) of time, then the original MAT scheme applies verbatim and $4/3$ is also the optimal DoF under this alternating CSIT model with a relaxed CSIT assumption. Therefore, the schemes $S^{4/3}_{3}$ and $S^{4/3}_{4}$ also achieve a DoF of $4/3$.
\end{itemize}

\subsection{Schemes achieving $3/2$ DoF}
The following schemes achieve $3/2$ DoF:
\begin{itemize}
\item $S^{3/2}_{1}$: using PD, NN for fractions $\left(\frac{1}{2}, \frac{1}{2}\right)$ and achieving $(d_{1},d_{2})=\left(1,\frac{1}{2}\right)$.

To show the achievability of $(d_{1},d_{2})=(1,\frac{1}{2})$, we show that it is possible to reliably transmit two symbols $(u_{1},u_{2})$ to receiver $1$ and 
one symbol $v$ to receiver $2$ in two channel uses. The CSIT configuration is chosen as PD at $t=1$ and NN at $t=2$.
At $t=1$, the encoder sends
\begin{align}
X(1)=\begin{bmatrix}u_{1}\\ u_{2}\end{bmatrix}+B\begin{bmatrix} v\\ 0\end{bmatrix}
\end{align}
where the precoding matrix $B$ is chosen such that $H(1)B=0$. The outputs at the receivers at $t=1$ are given as
\begin{align}
Y(1)&=H(1)\begin{bmatrix}u_{1}\\ u_{2}\end{bmatrix}, \\
&\triangleq L_{1}(u_{1},u_{2})\\
Z(1)&= G(1)\begin{bmatrix}u_{1}\\ u_{2}\end{bmatrix}+G(1)B\begin{bmatrix} v \\ 0 \end{bmatrix}\\
&\triangleq L_{2}(u_{1},u_{2}) +v.
\end{align}
\begin{figure}[t]
  \centering
\includegraphics[width=10.2cm]{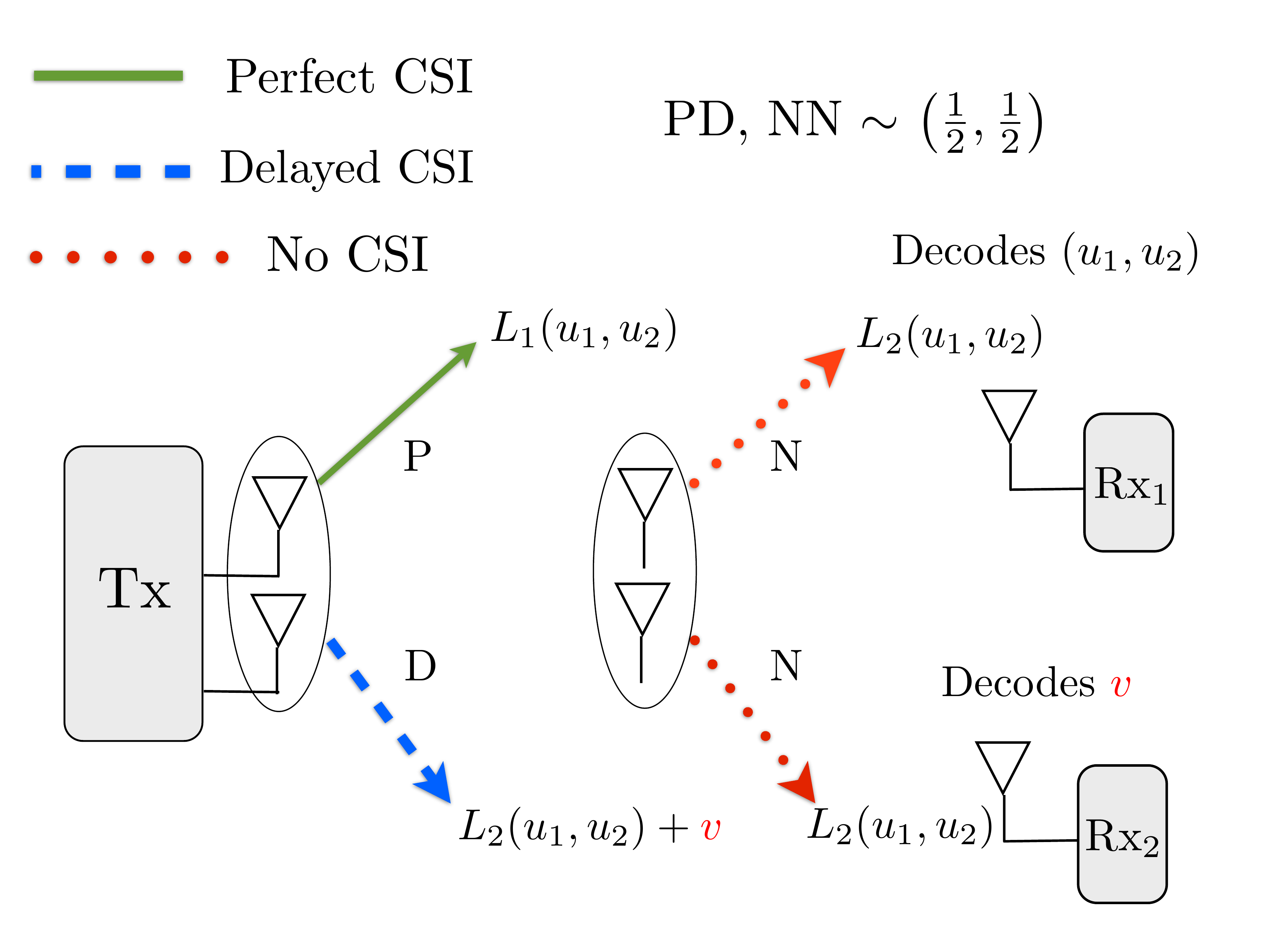}
\caption{Achieving $3/2$ DoF with (PD, NN) $\sim$ $(1/2, 1/2)$.}\label{Figure3by2PDNN}
\end{figure}
Due to delayed CSIT, the transmitter has access to $L_{2}(u_{1},u_{2})$. Hence, at $t=2$, it simply sends
\begin{align}
X(2)=\begin{bmatrix}L_{2}(u_{1},u_{2})\\0\end{bmatrix},
\end{align}
so that
\begin{align}
Y(2)= H(2)\begin{bmatrix}L_{2}(u_{1},u_{2})\\0\end{bmatrix}, \quad Z(2)= G(2)\begin{bmatrix}L_{2}(u_{1},u_{2})\\0\end{bmatrix}.
\end{align}
Having access to $L_{1}(u_{1},u_{2})$, along with $L_{2}(u_{1},u_{2})$, the symbols $(u_{1},u_{2})$ can be decoded at receiver $1$.  At receiver $2$, the symbol $v$ can be decoded from $Z(1)= L_{2}(u_{1},u_{2})+v$ by canceling out the interference $L_{2}(u_{1},u_{2})$ which is received at $t=2$. The scheme is illustrated in Figure \ref{Figure3by2PDNN}.

\item $S^{3/2}_{2}$: using DP, NN for fractions $\left(\frac{1}{2}, \frac{1}{2}\right)$ and achieving $(d_{1},d_{2})=\left(\frac{1}{2},1\right)$.

\item $S^{3/2}_{3}$: using PN, NP for fractions $\left(\frac{1}{2}, \frac{1}{2}\right)$ and achieving $(d_{1},d_{2})=\left(1,\frac{1}{2}\right)$.

To show the achievability of $(d_{1},d_{2})=(1,\frac{1}{2})$, we show that it is possible to reliably transmit two symbols $(u_{1},u_{2})$ to receiver $1$ and 
one symbol $v$ to receiver $2$ in two channel uses. The CSIT configuration is chosen as PN at $t=1$ and NP at $t=2$.
At $t=1$, the encoder sends
\begin{align}
X(1)=\begin{bmatrix}u_{1}\\ 0\end{bmatrix}+B(1)\begin{bmatrix} v\\ 0\end{bmatrix}
\end{align}
where the precoding matrix $B(1)$ is chosen such that $H(1)B(1)=0$. The outputs at receivers at $t=1$ are given as
\begin{align}
Y(1)&=H(1)\begin{bmatrix}u_{1}\\ 0\end{bmatrix}, \\
&\triangleq u_{1}\\
Z(1)&= G(1)\begin{bmatrix}u_{1}\\ 0\end{bmatrix}+G(1)B(1)\begin{bmatrix} v \\ 0 \end{bmatrix}\\
&\triangleq L(u_{1},v).
\end{align}
\begin{figure}[t]
  \centering
\includegraphics[width=10.2cm]{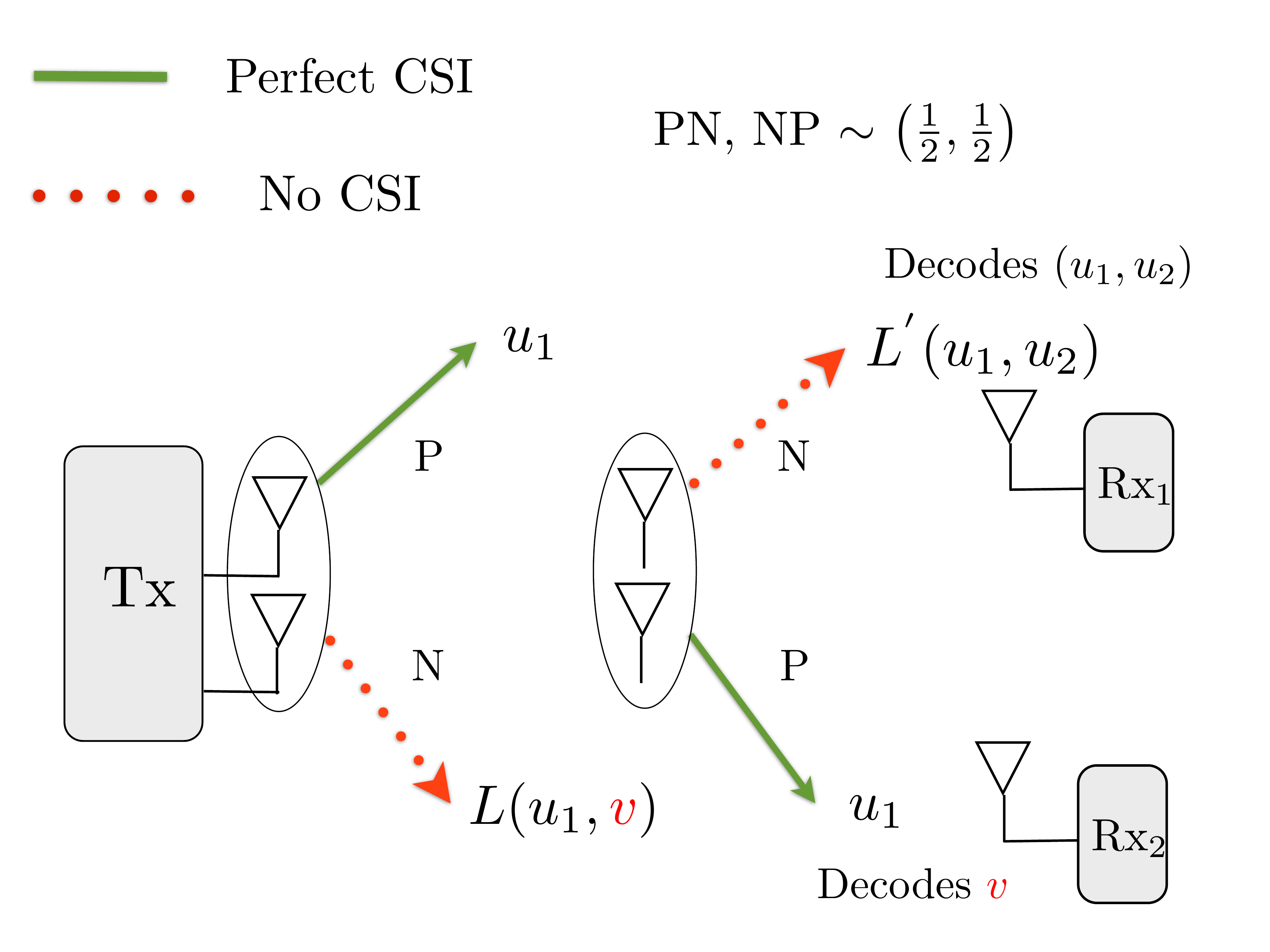}
\caption{Achieving $3/2$ DoF with (PN, NP) $\sim$ $(1/2, 1/2)$.}\label{Figure3by2PNNP}
\end{figure}
At this point, receiver $2$ requires $u_{1}$ cleanly in order to decode $v$. At $t=2$, the CSIT configuration changes to NP, and the transmitter can send $u_{1}$ cleanly to receiver $2$; but at the same time it
uses the second antenna to transmit $u_{2}$ which is intended for receiver $1$.
\begin{align}
X(2)=\begin{bmatrix}u_{1}\\ 0\end{bmatrix}+B(2)\begin{bmatrix} u_{2}\\ 0\end{bmatrix}
\end{align}
where the precoding matrix $B(2)$ is chosen such that $G(2)B(2)=0$
so that
\begin{align}
Y(2)&= H(2)\begin{bmatrix}u_{1}\\0\end{bmatrix} + H(2)B(2)\begin{bmatrix} u_{2}\\ 0\end{bmatrix}\\
&\triangleq L^{'}(u_{1},u_{2}),\\
Z(2)&= G(2)\begin{bmatrix}u_{1}\\0\end{bmatrix}+ G(2)B(2)\begin{bmatrix} u_{2}\\ 0\end{bmatrix}\\
&= G(2)\begin{bmatrix}u_{1}\\0\end{bmatrix}\triangleq u_{1}.
\end{align}
Having access to $u_{1}$, along with $L^{'}(u_{1},u_{2})$, the symbols $(u_{1},u_{2})$ can be decoded at receiver $1$.  At receiver $2$, the symbol $v$ can be decoded from $Z(1)= L(u_{1},v)$ by canceling out the interference $u_{1}$ which is received within noise distortion at $t=2$. The scheme is illustrated in Figure \ref{Figure3by2PNNP}.

\item $S^{3/2}_{4}$: using PN, NP for fractions $\left(\frac{1}{2}, \frac{1}{2}\right)$ and achieving $(d_{1},d_{2})=\left(\frac{1}{2},1\right)$.

\item $S^{3/2}_{5}$: using ND, PN for fractions $\left(\frac{1}{2}, \frac{1}{2}\right)$ and achieving $(d_{1},d_{2})=\left(1,\frac{1}{2}\right)$.

To show the achievability of $(d_{1},d_{2})=(1,\frac{1}{2})$, we show that it is possible to reliably transmit two symbols $(u_{1},u_{2})$ to receiver $1$ and 
one symbol $v$ to receiver $2$ in two channel uses. The CSIT configuration is chosen as ND at $t=1$ and PN at $t=2$.
At $t=1$, the encoder sends
\begin{align}
X(1)=\begin{bmatrix}u_{1}\\ u_{2}\end{bmatrix}
\end{align}
The outputs at receivers at $t=1$ are given as
\begin{align}
Y(1)&=H(1)\begin{bmatrix}u_{1}\\ u_{2}\end{bmatrix} \triangleq L_{1}(u_{1},u_{2}),\quad  Z(1)= G(1)\begin{bmatrix}u_{1}\\ u_{2}\end{bmatrix}\triangleq L_{2}(u_{1},u_{2}).
\end{align}
\begin{figure}[t]
  \centering
\includegraphics[width=10.2cm]{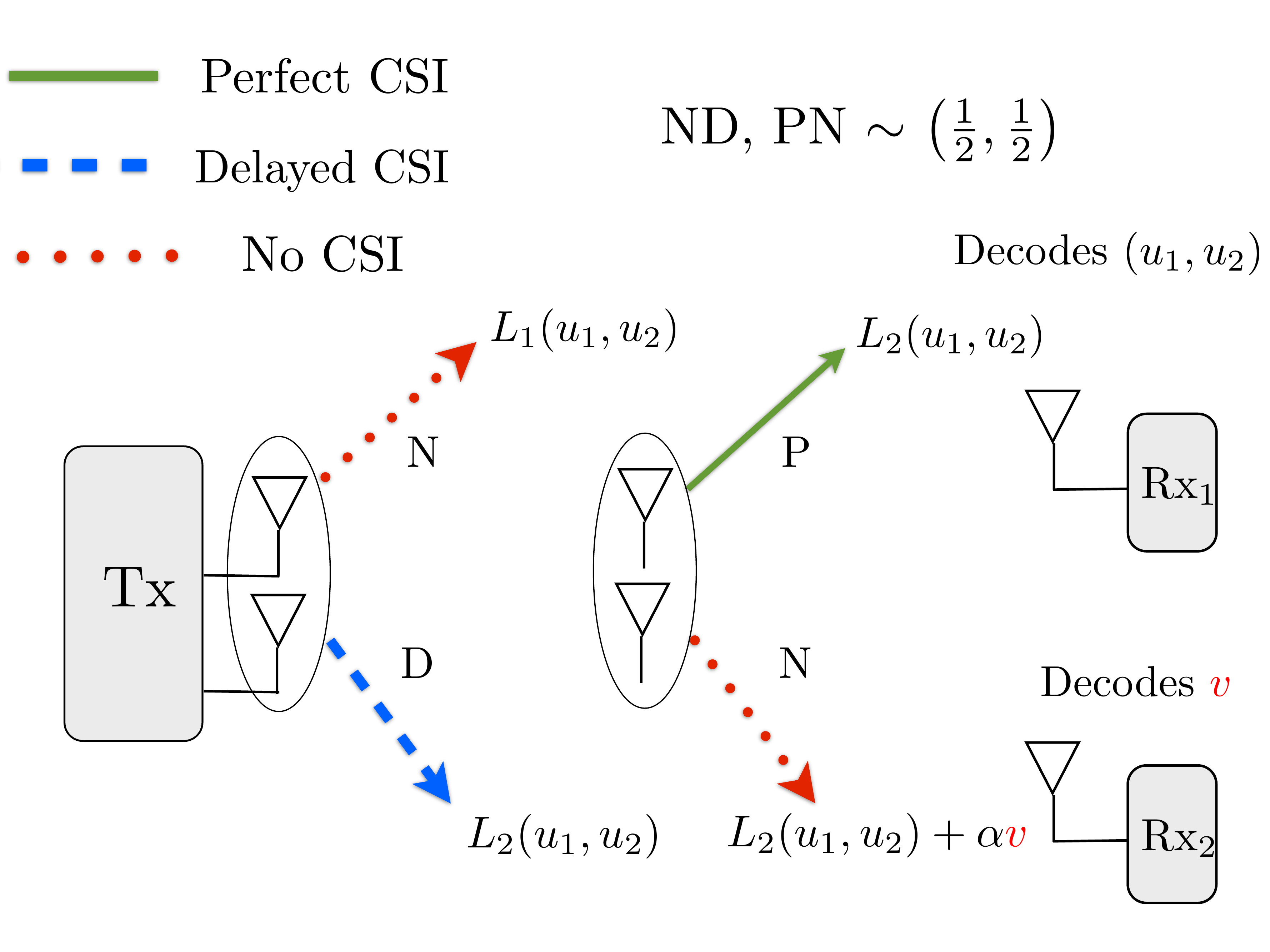}
\caption{Achieving $3/2$ DoF with (ND, PN) $\sim$ $(1/2, 1/2)$.}\label{Figure3by2PNND}
\end{figure}
At this point, side information $L_{2}(u_{1},u_{2})$ is created at receiver $2$, and if receiver $1$ can obtain $L_{2}(u_{1},u_{2})$ cleanly, then it can decode $(u_{1},u_{2})$. 
Due to delayed CSIT from receiver $2$ after $t=1$, the transmitter can obtain $L_{2}(u_{1},u_{2})$ within noise distortion.

At $t=2$, the CSIT configuration changes to PN, and the transmitter can send $L_{2}(u_{1},u_{2})$ cleanly to receiver $2$; but at the same time it
uses the second antenna to transmit $v$ which is intended for receiver $2$.
\begin{align}
X(2)=\begin{bmatrix}L_{2}(u_{1},u_{2})\\ 0\end{bmatrix}+B(2)\begin{bmatrix} v\\ 0\end{bmatrix}
\end{align}
where the precoding matrix $B(2)$ is chosen such that $H(2)B(2)=0$
so that
\begin{align}
Y(2)&= H(2)\begin{bmatrix}L_{2}(u_{1},u_{2})\\0\end{bmatrix} + H(2)B(2)\begin{bmatrix} v\\ 0\end{bmatrix}\\
&\triangleq L_{2}(u_{1},u_{2}),\\
Z(2)&= G(2)\begin{bmatrix}L_{2}(u_{1},u_{2})\\0\end{bmatrix}+ G(2)B(2)\begin{bmatrix} v\\ 0\end{bmatrix}\\
&= L_{2}(u_{1},u_{2})+ \alpha v.
\end{align}
Having access to $L_{1}(u_{1},u_{2})$, along with $L_{2}(u_{1},u_{2})$, the symbols $(u_{1},u_{2})$ can be decoded at receiver $1$.  At receiver $2$, the symbol $v$ can be decoded from $Z(2)= L_{2}(u_{1},u_{2})+\alpha v$ by canceling out the interference $L_{2}(u_{1},u_{2})$ which was received within noise distortion previously at $t=1$. The scheme is illustrated in Figure \ref{Figure3by2PNND}.

\item $S^{3/2}_{6}$: using DN, NP for fractions $\left(\frac{1}{2}, \frac{1}{2}\right)$ and achieving $(d_{1},d_{2})=\left(\frac{1}{2},1\right)$.
\end{itemize}

\subsection{Schemes achieving $5/3$ DoF}
The following schemes achieve $5/3$ DoF:
\begin{itemize}
\item $S^{5/3}_{1}$: using PD, DP for fractions $\left(\frac{2}{3},\frac{1}{3}\right)$ and achieving $(d_{1},d_{2})=\left(1,\frac{2}{3}\right)$.
\item $S^{5/3}_{2}$: using DP, PD for fractions $\left(\frac{1}{3},\frac{2}{3}\right)$ and achieving $(d_{1},d_{2})=\left(\frac{2}{3},1\right)$.
\item $S^{5/3}_{3}$: using PD, PN, NP for fractions $\left(\frac{1}{3},\frac{1}{3}, \frac{1}{3}\right)$ and achieving $(d_{1},d_{2})=\left(1,\frac{2}{3}\right)$.

In this scheme, we show that it is possible to reliably transmit three symbols $(u_{1},u_{2},u_{3})$ to receiver $1$ and two symbols $(v_{1},v_{2})$ to receiver $2$ in a total of three channel uses.
The CSIT states are chosen as PD at $t=1$,  PN at $t=2$, and NP at $t=3$.
At $t=1$, the encoder sends
\begin{align}
X(1)=\begin{bmatrix}u_{1} \\u_{2}\end{bmatrix} +B(1)\begin{bmatrix}v_{1} \\0\end{bmatrix},
\end{align}
where the precoding matrix $B(1)$ is chosen to satisfy $H(1)B(1)=0$. The channel outputs are given as
\begin{align}
Y(1)&=H(1)\begin{bmatrix}u_{1}\\ u_{2}\end{bmatrix}\\
&\triangleq L_{1}(u_{1},u_{2}),\\
Z(1)&= G(1)\begin{bmatrix}u_{1}\\ u_{2}\end{bmatrix} +G(1)B(1)\begin{bmatrix}v_{1}\\ 0\end{bmatrix}\\
&\triangleq L_{2}(u_{1},u_{2})+\alpha_{1}v_{1}.
\end{align}
Due to delayed CSIT, transmitter has access to $G(1)$ after $t=1$. It can reconstruct the interference $L_{2}(u_{1},u_{2})$ seen at receiver $2$.
Hence, at $t=2$, it sends
\begin{align}
X(2)=  \begin{bmatrix}L_{2}(u_{1},u_{2})\\ 0\end{bmatrix}+B(2)\begin{bmatrix}v_{2}\\ 0\end{bmatrix},
\end{align}
where the precoding matrix $B(2)$ is chosen to satisfy $H(2)B(2)=0$. The channel outputs are given as
\begin{align}
Y(2)&=H(2)\begin{bmatrix}L_{2}(u_{1},u_{2})\\ 0\end{bmatrix}\triangleq L_{2}(u_{1},u_{2})\\
Z(2)&= G(2)\begin{bmatrix}L_{2}(u_{1},u_{2})\\ 0\end{bmatrix}+B(2)\begin{bmatrix}v_{2}\\ 0\end{bmatrix}\\
&\triangleq L_{2}(u_{1},u_{2})+\alpha_{2}v_{2}.
\end{align}
The key consequence of this encoding step is that receiver $2$ still faces the \emph{same} interference (up to a known scaling factor) as it encountered at $t=1$.
However, to successfully decode $(v_{1},v_{2})$, it still requires this interference cleanly, i.e., it requires $L_{2}(u_{1},u_{2})$.

\begin{figure}[t]
  \centering
\includegraphics[width=11.5cm]{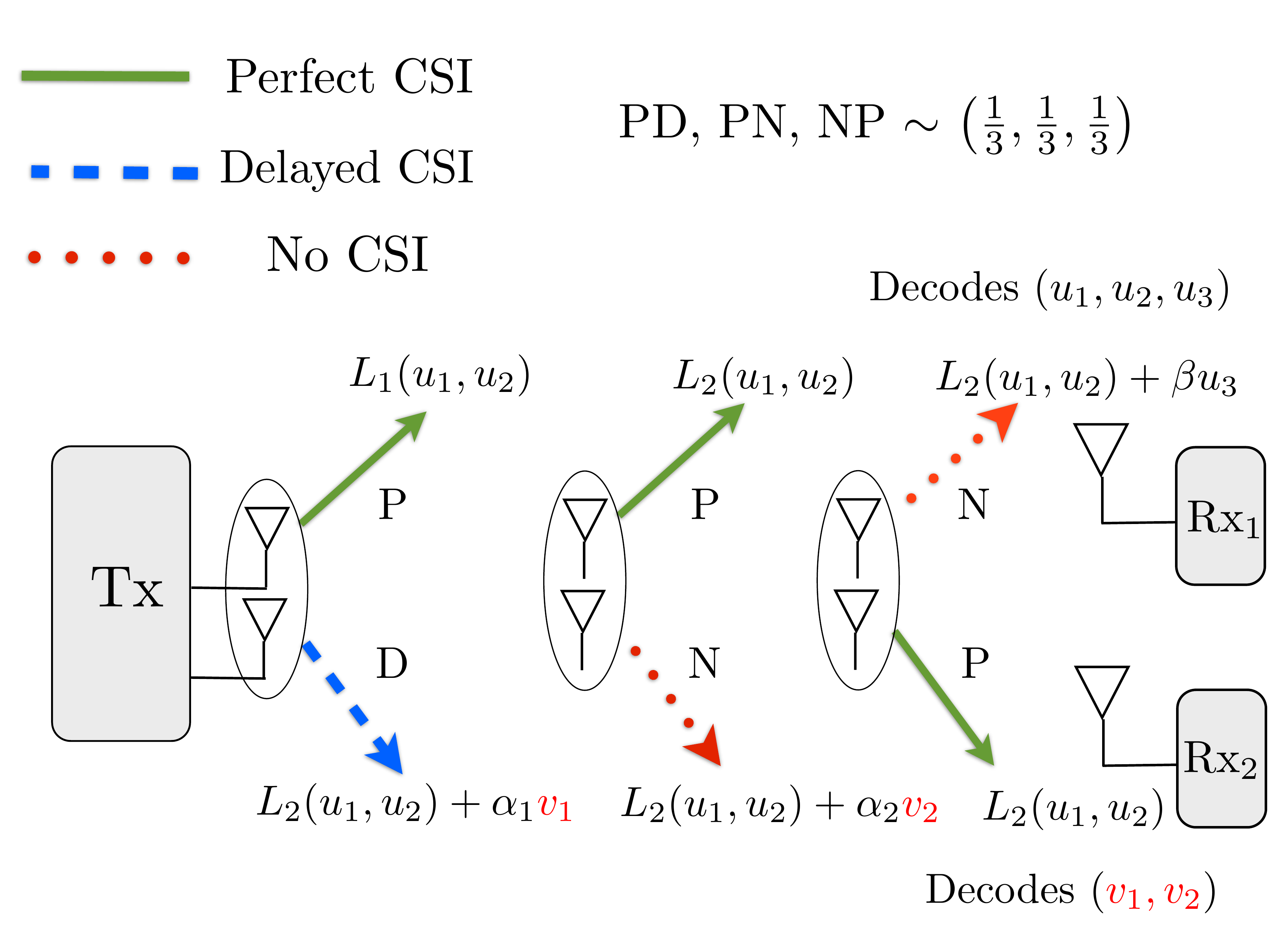}
\caption{Achieving $5/3$ DoF with (PD, PN, NP) $\sim$ $(1/3, 1/3,1/3)$.}\label{Figure5by3PDPNNP}
\end{figure}

The transmitter now uses the freedom provided under the \emph{alternating} CSIT model and switches from CSIT state PN at $t=2$ to the state NP at $t=3$.
Having access to $G(3)$, it sends
\begin{align}
X(3)= \begin{bmatrix}L_{2}(u_{1},u_{2})\\ 0\end{bmatrix} +B(3)\begin{bmatrix}u_{3}\\ 0\end{bmatrix},
\end{align}
where $G(3)B(3)=0$. The outputs are given as
\begin{align}
Y(3)&= H(3)\begin{bmatrix}L_{2}(u_{1},u_{2})\\ 0\end{bmatrix} +H(3)B(3)\begin{bmatrix}u_{3}\\ 0\end{bmatrix}\\
&\triangleq L_{2}(u_{1},u_{2})+\beta u_{3},\\
Z(3)&=G(3)\begin{bmatrix}L_{2}(u_{1},u_{2})\\ 0\end{bmatrix}\\
&= L_{2}(u_{1},u_{2}).
\end{align}
Having access to $(Y(1),Y(2),Y(3))$, the symbols $(u_{1},u_{2},u_{3})$ can be decoded. Finally, upon receiving $Z(3)$, receiver $2$ successfully decodes $(v_{1},v_{2})$.
The scheme is illustrated in Figure \ref{Figure5by3PDPNNP}.

Note that this scheme also shows that $5/3$ DoF is achievable as mentioned for schemes $S^{5/3}_{1}$ and $S^{5/3}_{2}$, since the states PD, DP at $t=2,3$ can always be 
used as PN, NP states as above by ignoring the respective delayed CSIT components.

\item $S^{5/3}_{4}$: using DP, PN, NP for fractions $\left(\frac{1}{3},\frac{1}{3}, \frac{1}{3}\right)$ and achieving $(d_{1},d_{2})=\left(\frac{2}{3},1\right)$.
\end{itemize}

\subsection{Scheme achieving $8/5$ DoF}
The following scheme achieves $8/5$ DoF:
\begin{itemize}
\item $S^{8/5}$: using DD, PN, NP for fractions $\left(\frac{1}{5},\frac{2}{5},\frac{2}{5}\right)$ and achieving $(d_{1},d_{2})=\left(\frac{4}{5},\frac{4}{5}\right)$.
\end{itemize}
To this end, we show that it is possible to reliably transmit $4$ symbols $(u_{1},u_{2},u_{3},u_{4})$ to receiver $1$, and $4$ symbols $(v_{1},v_{2},v_{3},v_{4})$ 
to receiver $2$ in a total of five channel uses. The CSIT configurations are chosen as DD, PN, NP, PN, and NP for $t=1,2,3,4$ and $t=5$ respectively.
At $t=1$, the transmitter sends the following:
\begin{align}
X(1)&=  \begin{bmatrix}u_{1}+v_{1}\\u_{2}+v_{2}\end{bmatrix},
\end{align}
so that the channel outputs are 
\begin{align}
Y(1)&=  H(1) \begin{bmatrix}u_{1}+v_{1}\\u_{2}+v_{2}\end{bmatrix}\\
&= A_{1}(u_{1},u_{2}) + B_{1}(v_{1},v_{2})\\
&\triangleq A_{1}+ B_{1},
\end{align}
and
\begin{align}
Z(1)&=  G(1) \begin{bmatrix}u_{1}+v_{1}\\u_{2}+v_{2}\end{bmatrix}\\
&= A_{2}(u_{1},u_{2}) + B_{2}(v_{1},v_{2})\\
&\triangleq A_{2}+ B_{2}.
\end{align}
Due to delayed CSIT from both receivers (the state DD at $t=1$), the transmitter can reconstruct $B_{1}$ and $A_{2}$ (which are the interference components at receivers $1$ and $2$ respectively).

At $t=2$, the transmitter sends $B_{1}$ cleanly to receiver $1$, and uses the second antenna to send $v_{3}$:
\begin{align}
X(2)= \begin{bmatrix}B_{1}\\ 0\end{bmatrix} +S(2)\begin{bmatrix}v_{3}\\ 0\end{bmatrix},
\end{align}
where $H(2)S(2)=0$. The outputs at $t=2$ are
\begin{align}
Y(2)&= H(2)\begin{bmatrix}B_{1}\\ 0\end{bmatrix} +H(2)S(2)\begin{bmatrix}v_{3}\\ 0\end{bmatrix}\triangleq B_{1}\\
Z(2)&= G(2)\begin{bmatrix}B_{1}\\ 0\end{bmatrix} +G(2)S(2)\begin{bmatrix}v_{3}\\ 0\end{bmatrix}\triangleq B_{1}+ B_{3}
\end{align}
where $B_{3}$ is a scaled version of $v_{3}$. 

At $t=3$, the transmitter switches the role by alternating to the NP  state and sends $A_{2}$ cleanly to receiver $2$ and uses the second antenna to send $u_{3}$. We thus have,
\begin{align}
Y(3)&= A_{2}+A_{3}, \quad Z(3)= A_{2}.
\end{align}

At this point, we observe that receiver $1$ requires $A_{2}$ and receiver $2$ requires $B_{1}$. Moreover, the only interference that receiver $1$ has seen so far is $B_{1}$;
and the only interference that receiver $2$ has encountered so far is $A_{2}$. 
\begin{figure}[t]
  \centering
\includegraphics[width=11.5cm]{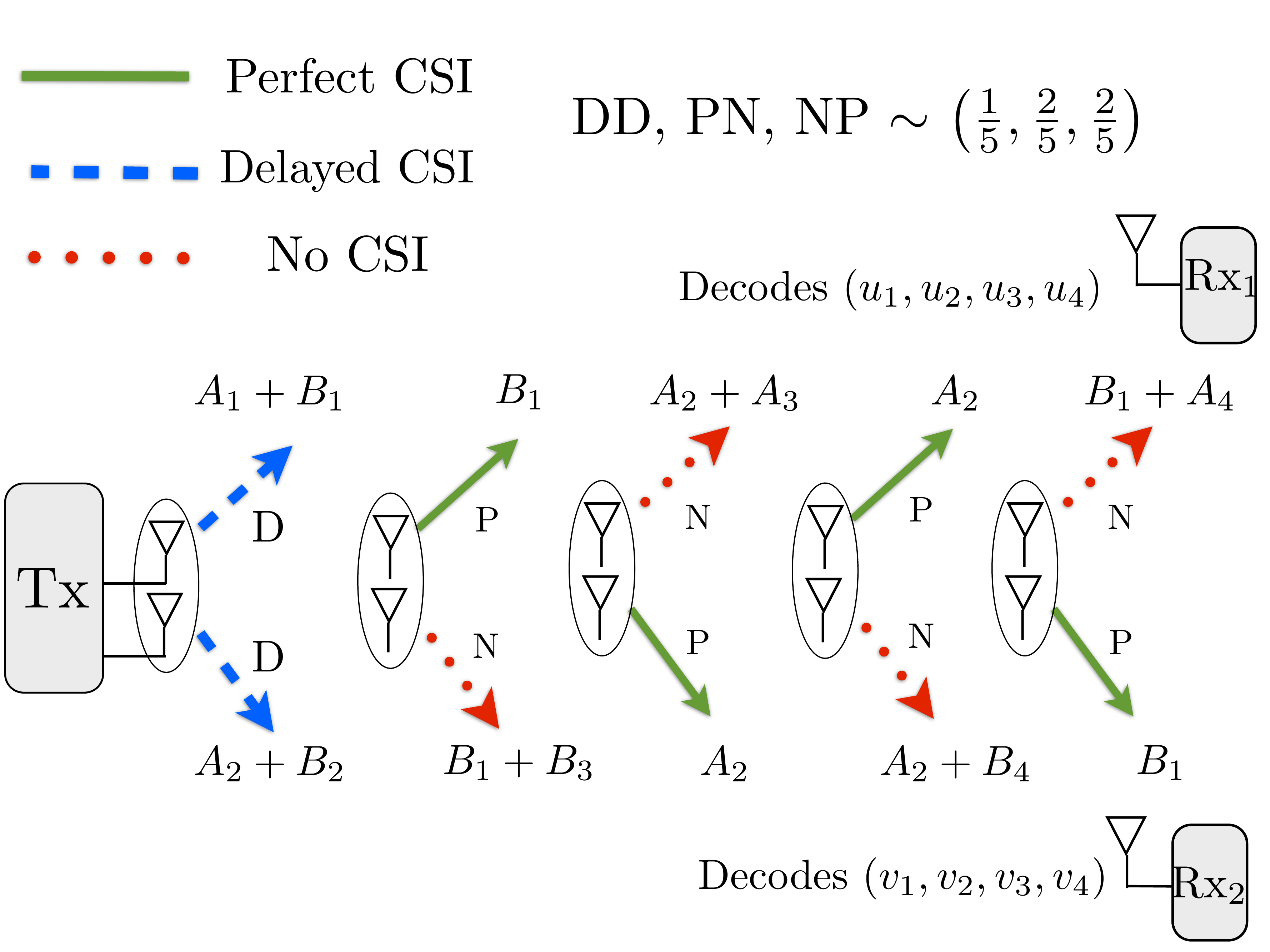}
\caption{Achieving $8/5$ DoF with (DD, PN, NP) $\sim$ $(1/5, 2/5, 2/5)$.}\label{Figure8by5}
\end{figure}

At $t=4$, transmitter is in the PN state and it sends $A_{2}$ cleanly to receiver $1$, and uses the second antenna to send $v_{4}$:
\begin{align}
X(4)= \begin{bmatrix}A_{2}\\ 0\end{bmatrix} +S(4)\begin{bmatrix}v_{4}\\ 0\end{bmatrix},
\end{align}
where $H(4)S(4)=0$. The outputs at $t=2$ are
\begin{align}
Y(4)&= H(4)\begin{bmatrix}A_{2}\\ 0\end{bmatrix} +H(4)S(4)\begin{bmatrix}v_{3}\\ 0\end{bmatrix}\triangleq A_{2}\\
Z(4)&= G(4)\begin{bmatrix}A_{2}\\ 0\end{bmatrix} +G(4)S(4)\begin{bmatrix}v_{3}\\ 0\end{bmatrix}\triangleq A_{2}+ B_{4}
\end{align}
where $B_{4}$ is a scaled version of $v_{4}$.
 
 At $t=3$, the transmitter switches the role by alternating to the NP  state and sends $B_{1}$ cleanly to receiver $2$ and uses the second antenna to send $u_{4}$. We thus have,
\begin{align}
Y(5)&= B_{1}+A_{4}, \quad Z(5)= B_{1}.
\end{align}
To summarize, the channel outputs can be written as
\begin{align}
\mathbf{Y}= \begin{bmatrix}A_{1}+B_{1}\\ B_{1}\\ A_{2}+A_{3}\\ A_{2}\\B_{1}+A_{4}\end{bmatrix}, \quad \mbox{and} \quad \mathbf{Z}&= \begin{bmatrix}A_{2}+B_{2}\\ B_{1}+B_{3}\\ A_{2}\\ A_{2}+B_{4}\\B_{1}\end{bmatrix},
\end{align}
and $(A_{1}(u_{1},u_{2}),A_{2}(u_{1},u_{2}),A_{3}(u_{3}),A_{4}(u_{4}))$ (and thus $(u_{1},u_{2},u_{3},u_{4})$) are decodable at receiver $1$; and similarly $(v_{1},\ldots,v_{4})$ are decoded at receiver $2$.
The scheme is illustrated in Figure \ref{Figure8by5}.
Thus, in order to achieve the DoF pair $(4/5, 4/5)$, interference can occupy at most one dimension in the five-dimensional output space at each receiver. This is precisely what  alternation allows the transmitter to accomplish by jointly using  DD, PN and NP states.  

\begin{remark}
We note here that if the CSIT state at time $t$ is modeled as an i.i.d. random variable, i.e., $CSIT(t)= I_{1}I_{2}$, with probability $\lambda_{I_{1}I_{2}}$, the corresponding DoF regions and claims
would continue to hold. For instance, consider the case in which the states DD, PN, NP are present for fractions $(\frac{1}{5},\frac{2}{5},\frac{2}{5})$ and scheme $S^{8/5}$ is shown to achieve $8/5$ DoF. The scheme presented above uses the state DD at $t=1$ and the states $PN, NP$ are used thereafter at $t=2,\ldots,5$. This scheme indicates that in order to achieve $8/5$ DoF,  the DD state should occur before the PN and NP states.  
Now, consider the case in which CSIT state is modeled as an i.i.d. random variable as follows:
\begin{align}
CSIT(t)=\begin{cases} 
DD &\mbox{w.p. } \frac{1}{5}\\ 
PN &\mbox{w.p. } \frac{2}{5}\\ 
NP &\mbox{w.p. } \frac{2}{5}.
\end{cases} 
\end{align}
To substantiate the claim that $8/5$ DoF is also achievable under this model, consider a long block of size $n$. By strong typicality,  as $n\rightarrow \infty$, $1/5$ of the total states would be DD states, $2/5$ would be NP states and $2/5$ would be PN states. Now consider a sequence of such blocks, indexed as $b=1,\ldots,B$. In any given block $b$, the transmitter would use the DD states from the previous block $(b-1)$ along with the PN, NP states from the current block $b$ as it does for scheme $S^{8/5}$. By letting $B \rightarrow \infty$, this block-Markov modification of the original constituent scheme takes care of causality issues,  and guarantees that the DoF claims would continue to hold if the CSIT state evolves in an i.i.d. manner over time.
\end{remark}

\section{Achieving $\mathcal{D}(\lambda)$}\label{Sec:Achievability}
We need to show the achievability of the DoF region
\begin{align}
d_{1}&\leq 1\label{PAlt1}\\
d_{2}&\leq 1\label{PAlt2}\\
d_{1}+2d_{2}&\leq 2+\lambda_{P}\label{PAlt3}\\
2d_{1}+d_{2}&\leq 2+\lambda_{P}\label{PAlt4}\\
d_{1}+d_{2}&\leq 1+ \lambda_{P}+ \lambda_{D}\label{PAlt5}.
\end{align}
We first note that the DoF region takes two different shapes, depending on whether the $(d_{1}+d_{2})$ upper bound in 
(\ref{PAlt5}) is active or not. We thus have two cases:
\begin{itemize}
\item Case A: $(d_{1}+d_{2})$ bound is not active. This corresponds to the following condition:
\begin{align}
\frac{2(2+\lambda_{P})}{3}&\leq 1+ \lambda_{P}+\lambda_{D}\nonumber,
\end{align}
which by using $\lambda_{P}+\lambda_{D}+\lambda_{N}=1$, simplifies to
\begin{figure}[t]
  \centering
\includegraphics[width=11.2cm]{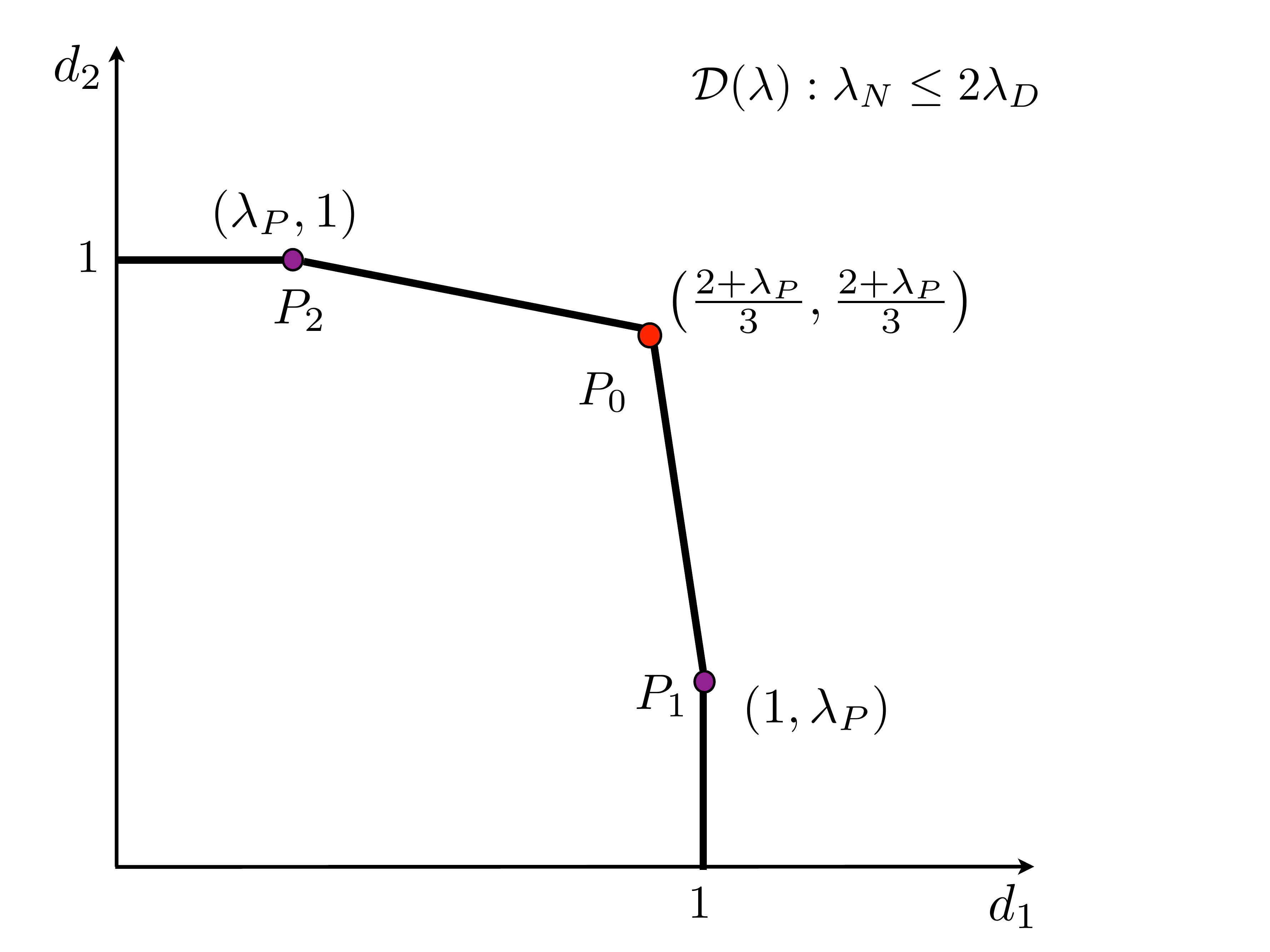}
\caption{DoF region for  Case A: $\lambda_{N}\leq 2\lambda_{D}$.}\label{Fig:CaseA}
\end{figure}
\begin{figure}[t]
  \centering
\includegraphics[width=11.2cm]{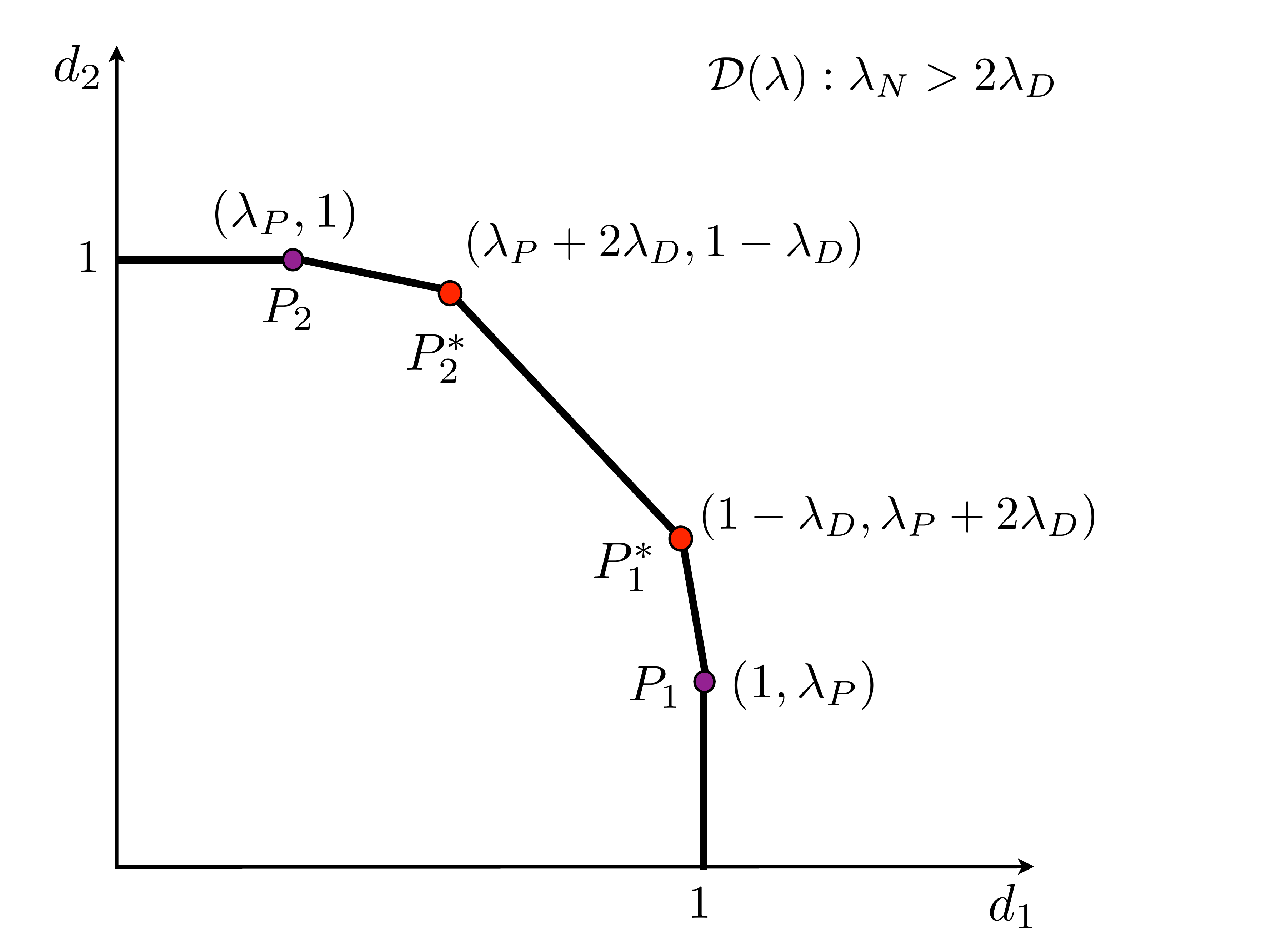}
\caption{DoF region for Case B: $\lambda_{N}> 2\lambda_{D}$.}\label{Fig:CaseB}
\end{figure}
\begin{align}
\lambda_{N}\leq 2\lambda_{D}.\label{ConditionA}
\end{align}
\item Case B: $(d_{1}+d_{2})$ bound is not active. This corresponds to the following condition:
\begin{align}
\lambda_{N}> 2\lambda_{D}.\label{ConditionB}
\end{align}
\end{itemize}
The DoF regions corresponding to the Cases A and B are shown in Figures \ref{Fig:CaseA} and \ref{Fig:CaseB} respectively. 
In both cases A and B, the corner points $P_{1}$ and $P_{2}$ remain fixed. We first show the achievability for $P_{1}$.
To this end, we show the achievability of the following pair:
\begin{align}
P_{1}: (d_{1},d_{2})= (1, \lambda_{P}).
\end{align}
This point can be achieved by the scheme in Table \ref{TableP1}.
\begin{table*}[h]
\begin{center}
\caption{Achieving $P_{1}$: $(d_{1},d_{2})=\left(1, \lambda_{PP}+\lambda_{PD}+\lambda_{PN}\right)$}\vspace{0.1in}
    \begin{tabular}{ | c | c | c|  c | c |}
    \hline
    Constituent Scheme              & CS-Fraction  & CS-$(d_{1},d_{2})$   	        &  Fraction                  &     Contribution to $(d_{1},d_{2})$\\\hline
        PP               &    	$1$          &    $(1,1)$	                  &     $\lambda_{PP}$     &      $(\lambda_{PP},\lambda_{PP})$\\\hline
   PD, DP  & $\left(\frac{1}{2},\frac{1}{2}\right)$  & $\left(1,\frac{1}{2}\right)$ &   $2\lambda_{PD}$    &      $(2\lambda_{PD}, \lambda_{PD})$\\\hline
   PN, NP  & $\left(\frac{1}{2},\frac{1}{2}\right)$ & $\left(1,\frac{1}{2}\right)$  &   $2\lambda_{PN}$   &      $(2\lambda_{PN}, \lambda_{PN})$\\\hline
   DN, ND & $\left(\frac{1}{2},\frac{1}{2}\right)$  & $(1,0)$ &   $2\lambda_{DN}$   &      $(2\lambda_{DN}, 0)$\\\hline
        DD      &  $1$      & $(1,0)$                                                       &   $\lambda_{DD}$     &	   $(\lambda_{DD}, 0)$\\\hline
        NN       & $1$       & $(1,0)$                                                              &   $\lambda_{NN}$     &      $(\lambda_{NN}, 0)$\\\hline
        $\sum$   &      &                                                         &      $1$                         &      $(1, \lambda_{P})$\\\hline

    \end{tabular}\label{TableP1}
\end{center}
\end{table*}
The achievability for the corner point $P_{2}$ follows due to symmetry with respect to $P_{1}$.
In the next sub-sections, we present the achievable schemes for Cases A and B.
\subsection{Achievability for Case A}
In this section, we show the achievability of $\mathcal{D}(\lambda)$ when $\lambda_{N}\leq 2\lambda_{D}$, which is equivalent to
\begin{align}
\lambda_{NN}+\lambda_{PN}&\leq 2\lambda_{DD} + 2\lambda_{PD} + \lambda_{DN}\label{EA}.
\end{align}
To this end, we sub-classify Case A into three mutually exclusive sub-cases as follows:
\begin{itemize}
\item Case A1: 
\begin{align}
\lambda_{NN}&\leq 2\lambda_{DD}\label{A1C1t}\\
\lambda_{PN}&\leq 2\lambda_{PD} + \lambda_{DN}\label{A1C2t}.
\end{align}
Note that (\ref{A1C1t})-(\ref{A1C2t}) imply that (\ref{EA}) is satisfied. 
\item Case A2: 
\begin{align}
\lambda_{NN}&\leq 2\lambda_{DD}\label{A2C1t}\\
\lambda_{PN}&> 2\lambda_{PD} + \lambda_{DN}\label{A2C2t}\\
\lambda_{NN}+\lambda_{PN}&\leq 2\lambda_{DD} + 2\lambda_{PD} + \lambda_{DN}\label{A2EAt}.
\end{align}
\item Case A3: 
\begin{align}
\lambda_{NN}&> 2\lambda_{DD}\label{A3C1t}\\
\lambda_{PN}&\leq 2\lambda_{PD} + \lambda_{DN}\label{A3C2t}\\
\lambda_{NN}+\lambda_{PN}&\leq 2\lambda_{DD} + 2\lambda_{PD} + \lambda_{DN}\label{A3EAt}.
\end{align}
\end{itemize}
\begin{remark}
Before proceeding, we give the intuition for the classification of Case A into the aforementioned three sub-cases.
By denoting
\begin{align}
&L_{1}\triangleq \lambda_{NN}, \qquad L_{2}\triangleq \lambda_{PN}\\
&R_{1}\triangleq 2\lambda_{DD}, \qquad R_{2}\triangleq  2\lambda_{PD} + \lambda_{DN},
\end{align}
the condition (\ref{EA}) can also be interpreted as follows:
\begin{align}
L_{1}+ L_{2}&\leq R_{1}+ R_{2}.
\end{align}
This inequality can be separately broken into pair-wise comparisons between the terms $(L_{1}, R_{1})$
and $(L_{2},R_{2})$. For instance, 
\begin{itemize}
\item Case A1 corresponds to $L_{1}\leq R_{1}, L_{2}\leq R_{2}$; 
\item Case A2 corresponds to $L_{1}\leq R_{1}, L_{2}>R_{2}, L_{1}+L_{2}\leq R_{1}+R_{2}$; 
\item Case A3 corresponds to
$L_{1}> R_{1}, L_{2}\leq R_{2}, L_{1}+L_{2}\leq R_{1}+R_{2}$.
\end{itemize}
\end{remark}

\subsubsection{Case A1}
This sub-case corresponds to the following conditions:
\begin{align}
\lambda_{NN}&\leq 2\lambda_{DD}\label{A1C1}\\
\lambda_{PN}&\leq 2\lambda_{PD} + \lambda_{DN}\label{A1C2}.
\end{align}
The condition (\ref{A1C1}) suggests that the state NN can be fully alternated with state DD using the scheme $S^{4/3}_{2}$.
The condition (\ref{A1C2}) suggests that the states (PN, NP) can be fully alternated with the states (PD, DP) using $(S^{5/3}_{3}, S^{5/3}_{4})$ 
and with (DN, ND) using schemes $(S^{3/2}_{5}, S^{3/2}_{6})$.

Inspired by these observations, in Table \ref{TableP0A1}, we present the scheme that achieves the DoF pair corresponding to $P_{0}$:
\begin{align}
(d_{1},d_{2})=\left(\frac{2+\lambda_{P}}{3},\frac{2+\lambda_{P}}{3}\right).
\end{align}
\begin{table*}[h]
\begin{center}
\caption{Case A1: achieving $P_{0}$: $(d_{1},d_{2})=\left(\frac{2+\lambda_{P}}{3},\frac{2+\lambda_{P}}{3}\right)$}\vspace{0.1in}
    \begin{tabular}{ | c | c | c|  c | c |}
    \hline
    Constituent Scheme (CS)              & CS-Fraction  				          & CS-$(d_{1},d_{2})$   	        &  Fraction                                  \\\hline
        PP       &                				$1$         			          &    $(1,1)$	                  		&     $\lambda_{PP}$              \\\hline
        
   PD, DP  &              $\left(\frac{2}{3},\frac{1}{3}\right)$           	          & $\left(1,\frac{2}{3}\right)$ 	&   $\lambda_{PD}-q_{1}$    \\\hline
   PD, DP  &              $\left(\frac{1}{3},\frac{2}{3}\right)$ 	          	 & $\left(\frac{2}{3},1\right)$ 	&   $\lambda_{PD}-q_{1}$    \\\hline
PD, NP, PN  &        $\left(\frac{1}{3},\frac{1}{3}, \frac{1}{3}\right)$       & $\left(1,\frac{2}{3}\right)$        &   $3q_{1}$         		    \\\hline
DP, NP, PN  &        $\left(\frac{1}{3},\frac{1}{3}, \frac{1}{3}\right)$       & $\left(\frac{2}{3},1\right)$        &   $3q_{1}$         		    \\\hline
 	
	PN, ND     &      $\left(\frac{1}{2},\frac{1}{2}\right)$           	          & $\left(1,\frac{1}{2}\right)$ 	&   $q_{2}$    \\\hline 
	NP, DN      &     $\left(\frac{1}{2},\frac{1}{2}\right)$           	          & $\left(\frac{1}{2},1\right)$ 	&   $q_{2}$    \\\hline

	DD        &             $1$						&  $\left(\frac{2}{3},\frac{2}{3}\right)$            &  $\lambda_{DD}- \frac{\lambda_{NN}}{2}$  \\\hline
DD, NN        & 	         $\left(\frac{1}{3},\frac{2}{3}\right)$      &  $\left(\frac{2}{3},\frac{2}{3}\right)$            &  $\frac{3\lambda_{NN}}{2}$  \\\hline
DN, ND, NN        & 	         $\left(\frac{1}{3},\frac{1}{3},\frac{1}{3}\right)$      &  $\left(\frac{2}{3},\frac{2}{3}\right)$            &  $2\lambda_{DN}-q_{2}$  \\\hline

           &      &                                                         &      $\sum = 1$                               \\\hline

    \end{tabular}\label{TableP0A1}
\end{center}
\end{table*}

The scheme in Table \ref{TableP0A1} works any $q_{1}, q_{2}$ satisfying the following three conditions:
\begin{align}
q_{1}&\leq \lambda_{PD} \label{PNA10}\\
q_{2}&\leq 2\lambda_{DN}\label{PNA11}\\
2q_{1}+ \frac{q_{2}}{2} &= \lambda_{PN}.\label{PNA1}
\end{align}
The conditions (\ref{PNA10}) and (\ref{PNA11}) ensure that the fractions of constituent schemes are non-negative; and condition (\ref{PNA1}) ensures that all fractions sum to $1$
and the marginals of the original states are as desired. Note that for these to simultaneously hold, we require (\ref{A1C2}). Furthermore, the condition (\ref{A1C1})
ensures that the fraction of scheme DD, i.e., $\lambda_{DD}- \frac{\lambda_{NN}}{2}$ is non-negative. 

The achievable $d_{1}= d_{2}$ (due to symmetry) for this scheme is as follows:
\begin{align}
d_{1}=d_{2}&=\lambda_{PP}+ \frac{5}{3}(\lambda_{PD}+2q_{1}) + \frac{3q_{2}}{2} + \frac{2}{3}(\lambda_{NN}+\lambda_{DD}+2\lambda_{DN}-q_{2})\\
&= \lambda_{PP}+ \frac{2(\lambda_{NN}+\lambda_{DD})+ 5\lambda_{PD}+4\lambda_{DN}}{3} + \frac{5}{3}\underbrace{(2q_{1}+\frac{q_{2}}{2})}_{= \lambda_{PN} }\\
&= \frac{2(\lambda_{PP}+\lambda_{NN}+\lambda_{DD}+2\lambda_{PD}+2\lambda_{PN}+2\lambda_{DN})+ (\lambda_{PP}+\lambda_{PD}+\lambda_{PN})}{3}\\
&= \frac{2+ (\lambda_{PP}+\lambda_{PD}+\lambda_{PN})}{3}\\
&= \frac{2+\lambda_{P}}{3}.
\end{align}

\subsubsection{Case A2}
This sub-case corresponds to the following conditions:
\begin{align}
\lambda_{NN}&\leq 2\lambda_{DD}\label{A2C1}\\
\lambda_{PN}&> 2\lambda_{PD} + \lambda_{DN}\label{A2C2}\\
\lambda_{NN}+\lambda_{PN}&\leq 2\lambda_{DD} + 2\lambda_{PD} + \lambda_{DN}\label{A2EA}.
\end{align}
The condition (\ref{A2C1}) suggests that the state NN can be fully alternated with state DD using the scheme $S^{4/3}_{2}$.
The condition (\ref{A2C2}) suggests that the states (PD, DP) 
and (DN, ND) can be fully alternated with (PN, NP) using schemes  using $(S^{5/3}_{3}, S^{5/3}_{4})$  and $(S^{3/2}_{5}, S^{3/2}_{6})$ respectively.
Finally, the condition (\ref{A2EA}) suggests that the remaining portion of (PN, NP) states can be alternated with the DD state by using the scheme $S^{8/5}$.
\begin{table*}[t]
\begin{center}
\caption{Case A2: achieving $P_{0}$: $(d_{1},d_{2})=\left(\frac{2+\lambda_{P}}{3},\frac{2+\lambda_{P}}{3}\right)$}\vspace{0.1in}
    \begin{tabular}{ | c | c | c|  c | c |}
    \hline
    Constituent Scheme (CS)              & CS-Fraction  				          & CS-$(d_{1},d_{2})$   	        &  Fraction                                  \\\hline
        PP       &                				$1$         			          &    $(1,1)$	                  		&     $\lambda_{PP}$              \\\hline
        
PD, NP, PN  &        $\left(\frac{1}{3},\frac{1}{3}, \frac{1}{3}\right)$       & $\left(1,\frac{2}{3}\right)$        &   $3\lambda_{PD}$         		    \\\hline
DP, NP, PN  &        $\left(\frac{1}{3},\frac{1}{3}, \frac{1}{3}\right)$       & $\left(\frac{2}{3},1\right)$        &   $3\lambda_{PD}$         		    \\\hline
 	
	PN, ND     &      $\left(\frac{1}{2},\frac{1}{2}\right)$           	          & $\left(1,\frac{1}{2}\right)$ 	&   $2\lambda_{DN}$    \\\hline 
	NP, DN      &     $\left(\frac{1}{2},\frac{1}{2}\right)$           	          & $\left(\frac{1}{2},1\right)$ 	&   $2\lambda_{DN}$    \\\hline 

	DD, PN, NP & $\left(\frac{1}{5}, \frac{2}{5}, \frac{2}{5}\right)$   & $\left(\frac{4}{5}, \frac{4}{5}\right)$  & $\frac{5}{2}(\lambda_{PN}-2\lambda_{PD}-\lambda_{DN})>0$  \\\hline

DD, NN        & 	         $\left(\frac{1}{3},\frac{2}{3}\right)$      &  $\left(\frac{2}{3},\frac{2}{3}\right)$            &  $\frac{3\lambda_{NN}}{2}$  \\\hline	
	DD        &             $1$						&  $\left(\frac{2}{3},\frac{2}{3}\right)$            &  $\lambda_{DD}- \frac{1}{2}(\lambda_{NN}+\lambda_{PN}-2\lambda_{PD}-\lambda_{DN})\geq 0$  \\\hline

           &      &                                                         &      $\sum = 1$                               \\\hline

    \end{tabular}\label{TableP0A2}
\end{center}
\end{table*}

Inspired by these observations, in Table \ref{TableP0A2}, we present the scheme that achieves the DoF pair corresponding to $P_{0}$:
\begin{align}
(d_{1},d_{2})=\left(\frac{2+\lambda_{P}}{3},\frac{2+\lambda_{P}}{3}\right).
\end{align}
Note that the conditions (\ref{A2C2})-(\ref{A2EA}) ensure that the fractions of all constituent schemes are non-negative.
The achievable $d_{1}= d_{2}$ (due to symmetry) for this scheme is as follows:
\begin{align}
d_{1}=d_{2}&=\lambda_{PP}+ 5\lambda_{PD} + 3\lambda_{DN} + \frac{5}{3}(\lambda_{PN}-2\lambda_{PD}-\lambda_{DN})+ \frac{2(\lambda_{DD}+\lambda_{NN})}{3}\\
&= \frac{2+ (\lambda_{PP}+\lambda_{PD}+\lambda_{PN})}{3}\\
&= \frac{2+\lambda_{P}}{3}.
\end{align}

\subsubsection{Case A3}
This sub-case corresponds to the following conditions:
\begin{align}
\lambda_{NN}&> 2\lambda_{DD}\label{A3C1}\\
\lambda_{PN}&\leq 2\lambda_{PD} + \lambda_{DN}\label{A3C2}\\
\lambda_{NN}+\lambda_{PN}&\leq 2\lambda_{DD} + 2\lambda_{PD} + \lambda_{DN}\label{A3EA}.
\end{align}
The condition (\ref{A3C1}) suggests that the state DD can be fully alternated with state NN using the scheme $S^{4/3}_{2}$.
The condition (\ref{A3C2}) suggests that the states (PN, NP) can be fully alternated with the states (PD, DP) using $(S^{5/3}_{3}, S^{5/3}_{4})$ 
and with (DN, ND) using schemes $(S^{3/2}_{5}, S^{3/2}_{6})$. Finally, the condition (\ref{A3EA}) suggests that the remaining 
portion of the state NN can be alternated with the set of states (PD, DP) and (DN, ND). 

In Table \ref{TableP0A3}, we present the scheme that achieves the DoF pair corresponding to $P_{0}$:
\begin{align}
(d_{1},d_{2})=\left(\frac{2+\lambda_{P}}{3},\frac{2+\lambda_{P}}{3}\right)
\end{align}

\begin{table*}[h]
\begin{center}
\caption{Case A3: achieving $P_{0}$: $(d_{1},d_{2})=\left(\frac{2+\lambda_{P}}{3},\frac{2+\lambda_{P}}{3}\right)$}\vspace{0.1in}
    \begin{tabular}{ | c | c | c|  c | c |}
    \hline
    Constituent Scheme (CS)              & CS-Fraction  				          & CS-$(d_{1},d_{2})$   	        &  Fraction                                  \\\hline
        PP       &                				$1$         			          &    $(1,1)$	                  		&     $\lambda_{PP}$              \\\hline
        
   PD, DP  &              $\left(\frac{2}{3},\frac{1}{3}\right)$           	          & $\left(1,\frac{2}{3}\right)$ 	&   $\lambda_{PD}-q_{1}- \frac{q_{3}}{2}$    \\\hline
   PD, DP  &              $\left(\frac{1}{3},\frac{2}{3}\right)$ 	          	 & $\left(\frac{2}{3},1\right)$ 	&   $\lambda_{PD}-q_{1}- \frac{q_{3}}{2}$    \\\hline
PD, NP, PN  &        $\left(\frac{1}{3},\frac{1}{3}, \frac{1}{3}\right)$       & $\left(1,\frac{2}{3}\right)$        &   $3q_{1}$         		    \\\hline
DP, NP, PN  &        $\left(\frac{1}{3},\frac{1}{3}, \frac{1}{3}\right)$       & $\left(\frac{2}{3},1\right)$        &   $3q_{1}$         		    \\\hline

	PD, NN     &      $\left(\frac{1}{2},\frac{1}{2}\right)$           	          & $\left(1,\frac{1}{2}\right)$ 	&   $q_{3}$    \\\hline 
	DP, NN      &     $\left(\frac{1}{2},\frac{1}{2}\right)$           	          & $\left(\frac{1}{2},1\right)$ 	&   $q_{3}$    \\\hline 
 	
	PN, ND     &      $\left(\frac{1}{2},\frac{1}{2}\right)$           	          & $\left(1,\frac{1}{2}\right)$ 	&   $q_{2}$    \\\hline 
	NP, DN      &     $\left(\frac{1}{2},\frac{1}{2}\right)$           	          & $\left(\frac{1}{2},1\right)$ 	&   $q_{2}$    \\\hline

DD, NN        & 	         $\left(\frac{1}{3},\frac{2}{3}\right)$      &  $\left(\frac{2}{3},\frac{2}{3}\right)$            &  $3\lambda_{DD}$  \\\hline
DN, ND, NN        & 	         $\left(\frac{1}{3},\frac{1}{3},\frac{1}{3}\right)$      &  $\left(\frac{2}{3},\frac{2}{3}\right)$            &  $3q_{4}$  \\\hline
DN, ND        & 	         $\left(\frac{1}{2},\frac{1}{2}\right)$      &  $\left(\frac{2}{3},\frac{2}{3}\right)$            &  $2\lambda_{DN}-q_{2}-2q_{4}$  \\\hline

           &      &                                                         &      $\sum = 1$                               \\\hline

    \end{tabular}\label{TableP0A3}
\end{center}
\end{table*}

The scheme in Table \ref{TableP0A3} works any $(q_{1}, q_{2}, q_{3}, q_{4})$ satisfying the following conditions:
\begin{align}
q_{1}+\frac{q_{3}}{2}&\leq \lambda_{PD} \label{PNA30}\\
q_{2}+2q_{4}&\leq 2\lambda_{DN}\label{PNA31}\\
2q_{1}+ \frac{q_{2}}{2} &= \lambda_{PN}\label{PNA32}\\
q_{3}+q_{4}&= \lambda_{NN}-2\lambda_{DD}.\label{PNA33}
\end{align}
The conditions (\ref{PNA30})-(\ref{PNA31}) ensure that the fractions of constituent schemes are non-negative; and conditions (\ref{PNA32})-(\ref{PNA33}) ensure that all fractions sum to $1$
and the marginals of the original states are as desired. Note that for these to hold simultaneously, we require (\ref{A3C1})-(\ref{A3EA}). 

The achievable $d_{1}= d_{2}$ (due to symmetry) for this scheme are as follows:
\begin{align}
d_{1}=d_{2}&=\lambda_{PP}+ \frac{5}{3}\lambda_{PD} + 2\lambda_{DD} + 4\lambda_{DN} + \frac{5}{3}\underbrace{\left(2q_{1}+\frac{q_{2}}{2}\right)}_{=\lambda_{PN}} + \frac{2}{3}\underbrace{\left(q_{3}+q_{4}\right)}_{= \lambda_{NN}-2\lambda_{DD}}\\
&= \frac{2(\lambda_{PP}+\lambda_{NN}+\lambda_{DD}+2\lambda_{PD}+2\lambda_{PN}+2\lambda_{DN})+ (\lambda_{PP}+\lambda_{PD}+\lambda_{PN})}{3}\\
&= \frac{2+ (\lambda_{PP}+\lambda_{PD}+\lambda_{PN})}{3}\\
&= \frac{2+\lambda_{P}}{3}.
\end{align}
This completes the proof of achievability of $\mathcal{D}(\lambda)$ for Case A.

\subsection{Achievability for Case B}
In this section, we show the achievability of $\mathcal{D}(\lambda)$ when $\lambda_{N}> 2\lambda_{D}$, which is equivalent to
\begin{align}
\lambda_{NN}+\lambda_{PN}&> 2\lambda_{DD} + 2\lambda_{PD} + \lambda_{DN}\label{EB}.
\end{align}
Similar to Case A, we sub-classify Case B into three mutually exclusive sub-cases as follows:
\begin{itemize}
\item Case B1: 
\begin{align}
\lambda_{NN}&> 2\lambda_{DD}\label{B1C1t}\\
\lambda_{PN}&> 2\lambda_{PD} + \lambda_{DN}\label{B1C2t}.
\end{align}
Note that (\ref{B1C1t})-(\ref{B1C2t}) imply that (\ref{EB}) is satisfied. 
\item Case B2: 
\begin{align}
\lambda_{NN}&\leq 2\lambda_{DD}\label{B2C1t}\\
\lambda_{PN}&> 2\lambda_{PD} + \lambda_{DN}\label{B2C2t}\\
\lambda_{NN}+\lambda_{PN}&> 2\lambda_{DD} + 2\lambda_{PD} + \lambda_{DN}\label{B2EAt}.
\end{align}
\item Case B3: 
\begin{align}
\lambda_{NN}&> 2\lambda_{DD}\label{B3C1t}\\
\lambda_{PN}&\leq 2\lambda_{PD} + \lambda_{DN}\label{B3C2t}\\
\lambda_{NN}+\lambda_{PN}&> 2\lambda_{DD} + 2\lambda_{PD} + \lambda_{DN}\label{B3EAt}.
\end{align}
\end{itemize}
Here, we focus on the achievability for the corner point $P_{1}^{*}$:
\begin{align}
(d_{1},d_{2})&=(1-\lambda_{D}, \lambda_{P}+2\lambda_{D})\\
&= (1-\lambda_{DD}-\lambda_{PD}-\lambda_{DN}, \lambda_{PP}+2\lambda_{DD}+ 3\lambda_{PD}+2\lambda_{DN}+\lambda_{PN}).
\end{align}

\subsubsection{Case B1}
This sub-case corresponds to the following conditions:
\begin{align}
\lambda_{NN}&> 2\lambda_{DD}\label{B1C1}\\
\lambda_{PN}&> 2\lambda_{PD} + \lambda_{DN}\label{B1C2}.
\end{align}
The condition (\ref{B1C1}) suggests that the state DD can be fully alternated with state NN using the scheme $S^{4/3}_{2}$.
The condition (\ref{B1C2}) suggests that the states (PD, DP)  and (DN, ND)  can be fully alternated with the states (PN, NP) using 
the schemes $(S^{5/3}_{3}, S^{5/3}_{4})$ and $(S^{3/2}_{5}, S^{3/2}_{6})$ respectively.
\begin{table*}[h]
\begin{center}
\caption{Case B1: achieving $P_{1}^{*}$}\vspace{0.1in}
    \begin{tabular}{ | c | c | c|  c | c |}
    \hline
    Constituent Scheme (CS)              & CS-Fraction  				          & CS-$(d_{1},d_{2})$   	        &  Fraction                                  \\\hline
        PP       &                				$1$         			          &    $(1,1)$	                  		&     $\lambda_{PP}$              \\\hline
        
PD, NP, PN  &        $\left(\frac{1}{3},\frac{1}{3}, \frac{1}{3}\right)$       & $\left(1,\frac{2}{3}\right)$        &   $3\lambda_{PD}$         		    \\\hline
DP, NP, PN  &        $\left(\frac{1}{3},\frac{1}{3}, \frac{1}{3}\right)$       & $\left(\frac{2}{3},1\right)$        &   $3\lambda_{PD}$         		    \\\hline
 	
	PN, ND     &      $\left(\frac{1}{2},\frac{1}{2}\right)$           	          & $\left(1,\frac{1}{2}\right)$ 	&   $2\lambda_{DN}$    \\\hline 
	NP, DN      &     $\left(\frac{1}{2},\frac{1}{2}\right)$           	          & $\left(\frac{1}{2},1\right)$ 	&   $2\lambda_{DN}$    \\\hline 
	PN, NP      &     $\left(\frac{1}{2},\frac{1}{2}\right)$           	          & $\left(1,\frac{1}{2}\right)$ 	&   $2(\lambda_{PN}-2\lambda_{PD}-\lambda_{DN})>0$    \\\hline

DD, NN        & 	         $\left(\frac{1}{3},\frac{2}{3}\right)$      &  $\left(\frac{2}{3},\frac{2}{3}\right)$            &  $3\lambda_{DD}$  \\\hline
         NN        &                                $1$				&   $(1,0)$							&    $\lambda_{NN}-2\lambda_{DD}>0$ \\\hline

           &      &                                                         &      $\sum = 1$                               \\\hline

    \end{tabular}\label{TableP1SB1}
\end{center}
\end{table*}
In Table \ref{TableP1SB1}, we present the scheme that achieves the DoF pair corresponding to $P_{1}^{*}$:
\begin{align}
(d_{1},d_{2})&= (1-\lambda_{DD}-\lambda_{PD}-\lambda_{DN}, \lambda_{PP} + 2\lambda_{DD}+ 3\lambda_{PD}+2\lambda_{DN}+\lambda_{PN}).
\end{align}
The achievable $(d_{1},d_{2})$ are as follows:
\begin{align}
d_{1}&= \lambda_{PP}+ 5\lambda_{PD}+ 3\lambda_{DN}+ 2(\lambda_{PN}-2\lambda_{PD}-\lambda_{DN})+ 2\lambda_{DD}+\lambda_{NN}-2\lambda_{DD}\\
&= \lambda_{PP}+\lambda_{NN}+\lambda_{DD}+ 2\lambda_{PD}+2\lambda_{PN}+2\lambda_{DN}-\lambda_{DD}-\lambda_{PD}-\lambda_{DN}\\
&= 1-\lambda_{DD}-\lambda_{PD}-\lambda_{DN},
\end{align}
and
\begin{align}
d_{2}&=\lambda_{PP}+ 5\lambda_{PD}+ 3\lambda_{DN}+ \lambda_{PN}-2\lambda_{PD}-\lambda_{DN}+2\lambda_{DD}\\
&= \lambda_{PP}+ 2\lambda_{DD}+ 3\lambda_{PD}+ 2\lambda_{DN}+ \lambda_{PN}.
\end{align}

\subsubsection{Case B2}
This sub-case corresponds to the following conditions:
\begin{align}
\lambda_{NN}&\leq 2\lambda_{DD}\label{B2C1}\\
\lambda_{PN}&> 2\lambda_{PD} + \lambda_{DN}\label{B2C2}\\
\lambda_{NN}+\lambda_{PN}&> 2\lambda_{DD} + 2\lambda_{PD} + \lambda_{DN}.\label{B2EA}
\end{align}
The condition (\ref{B2C1}) suggests that the state NN can be fully alternated with state DD using the scheme $S^{4/3}_{2}$.
The condition (\ref{B2C2}) suggests that the states (PD, DP)  and (DN, ND)  can be fully alternated with the states (PN, NP) using 
the schemes $(S^{5/3}_{3}, S^{5/3}_{4})$ and $(S^{3/2}_{5}, S^{3/2}_{6})$ respectively. Finally, condition (\ref{B2EA}) suggests that the remaining 
DD state can be alternated with (PN, NP) using the scheme $S^{8/5}$.

\begin{table*}[h]
\begin{center}
\caption{Case B2: achieving $P_{1}^{*}$}\vspace{0.1in}
    \begin{tabular}{ | c | c | c|  c | c |}
    \hline
    Constituent Scheme (CS)              & CS-Fraction  				          & CS-$(d_{1},d_{2})$   	        &  Fraction                                  \\\hline
        PP       &                				$1$         			          &    $(1,1)$	                  		&     $\lambda_{PP}$              \\\hline
        
PD, NP, PN  &        $\left(\frac{1}{3},\frac{1}{3}, \frac{1}{3}\right)$       & $\left(1,\frac{2}{3}\right)$        &   $3\lambda_{PD}$         		    \\\hline
DP, NP, PN  &        $\left(\frac{1}{3},\frac{1}{3}, \frac{1}{3}\right)$       & $\left(\frac{2}{3},1\right)$        &   $3\lambda_{PD}$         		    \\\hline
 	
	PN, ND     &      $\left(\frac{1}{2},\frac{1}{2}\right)$           	          & $\left(1,\frac{1}{2}\right)$ 	&   $2\lambda_{DN}$    \\\hline 
	NP, DN      &     $\left(\frac{1}{2},\frac{1}{2}\right)$           	          & $\left(\frac{1}{2},1\right)$ 	&   $2\lambda_{DN}$    \\\hline 
	
DD, NN        & 	         $\left(\frac{1}{3},\frac{2}{3}\right)$      &  $\left(\frac{2}{3},\frac{2}{3}\right)$            &  $\frac{3\lambda_{NN}}{2}$  \\\hline
DD, PN, NP  &       $\left(\frac{1}{5},\frac{2}{5},\frac{2}{5}\right)$      &  $\left(\frac{4}{5},\frac{4}{5}\right)$   &  $5\left(\lambda_{DD}-\frac{\lambda_{NN}}{2}\right)\geq 0$\\\hline

	PN, NP     &      $\left(\frac{1}{2},\frac{1}{2}\right)$           	          & $\left(1,\frac{1}{2}\right)$ 	&   $2\left(\lambda_{NN}+\lambda_{PN}-2\lambda_{DD}-2\lambda_{PD}-\lambda_{DN}\right)>0$    \\\hline 

           &      &                                                         &      $\sum = 1$                               \\\hline

    \end{tabular}\label{TableP1SB2}
\end{center}
\end{table*}
In  Table \ref{TableP1SB2}, we present the scheme that achieves the DoF pair corresponding to $P_{1}^{*}$:
\begin{align}
(d_{1},d_{2})&= (1-\lambda_{DD}-\lambda_{PD}-\lambda_{DN}, \lambda_{PP}+2\lambda_{DD}+ 3\lambda_{PD}+2\lambda_{DN}+\lambda_{PN}).
\end{align}
The achievable $(d_{1},d_{2})$ are as follows:
\begin{align}
d_{1}&= \lambda_{PP}+ 5\lambda_{PD}+ 3\lambda_{DN}+ \lambda_{NN}+2\lambda_{PN}-2\lambda_{DN}-4\lambda_{PD}\\
&= \lambda_{PP}+\lambda_{PD}+\lambda_{DN}+\lambda_{NN}+2\lambda_{PN}\\
&= 1-\lambda_{DD}-\lambda_{PD}-\lambda_{DN},
\end{align}
and
\begin{align}
d_{2}&=\lambda_{PP}+ 5\lambda_{PD}+ 3\lambda_{DN}+ \lambda_{PN}-2\lambda_{PD}-\lambda_{DN}+2\lambda_{DD}\\
&= \lambda_{PP}+ 2\lambda_{DD}+ 3\lambda_{PD}+ 2\lambda_{DN}+ \lambda_{PN}.
\end{align}

\subsubsection{Case B3}
This sub-case corresponds to the following conditions:
\begin{align}
\lambda_{NN}&> 2\lambda_{DD}\label{B3C1}\\
\lambda_{PN}&\leq 2\lambda_{PD} + \lambda_{DN}\label{B3C2}\\
\lambda_{NN}+\lambda_{PN}&> 2\lambda_{DD} + 2\lambda_{PD} + \lambda_{DN}.\label{B3EA}
\end{align}
The condition (\ref{B3C1}) suggests that the state DD can be fully alternated with state NN using the scheme $S^{4/3}_{2}$.
The condition (\ref{B3C2}) suggests that the states (PN, NP) can be fully alternated with (PD, DP)  and (DN, ND)  using 
the schemes $(S^{5/3}_{3}, S^{5/3}_{4})$ and $(S^{3/2}_{5}, S^{3/2}_{6})$ respectively. Finally, condition (\ref{B3EA}) suggests that the remaining 
fraction of (PD, DP) and (DN, ND) states can be alternated with the state NN.

\begin{table*}[h]
\begin{center}
\caption{Case B3: achieving $P_{1}^{*}$}\vspace{0.1in}
    \begin{tabular}{ | c | c | c|  c | c |}
    \hline
    Constituent Scheme (CS)              & CS-Fraction  				          & CS-$(d_{1},d_{2})$   	        &  Fraction                                  \\\hline
        PP       &                				$1$         			          &    $(1,1)$	                  		&     $\lambda_{PP}$              \\\hline
        
PD, NP, PN  &        $\left(\frac{1}{3},\frac{1}{3}, \frac{1}{3}\right)$       & $\left(1,\frac{2}{3}\right)$        &   $q_{1}$         		    \\\hline
DP, NP, PN  &        $\left(\frac{1}{3},\frac{1}{3}, \frac{1}{3}\right)$       & $\left(\frac{2}{3},1\right)$        &   $q_{1}$         		    \\\hline
 	
	PN, ND     &      $\left(\frac{1}{2},\frac{1}{2}\right)$           	          & $\left(1,\frac{1}{2}\right)$ 	&   $q_{2}$    \\\hline 
	NP, DN      &     $\left(\frac{1}{2},\frac{1}{2}\right)$           	          & $\left(\frac{1}{2},1\right)$ 	&   $q_{2}$    \\\hline 
	
	PD, NN     &      $\left(\frac{1}{2},\frac{1}{2}\right)$           	          & $\left(1,\frac{1}{2}\right)$ 	&   $2\left(\lambda_{PD}-\frac{q_{1}}{3}\right)$    \\\hline 
	DP, NN      &     $\left(\frac{1}{2},\frac{1}{2}\right)$           	          & $\left(\frac{1}{2},1\right)$ 	&   $2\left(\lambda_{PD}-\frac{q_{1}}{3}\right)$    \\\hline

DD, NN        & 	         $\left(\frac{1}{3},\frac{2}{3}\right)$      &  $\left(\frac{2}{3},\frac{2}{3}\right)$            &  $3\lambda_{DD}$  \\\hline
DN, ND, NN  &       $\left(\frac{1}{3},\frac{1}{3},\frac{1}{3}\right)$      &  $\left(\frac{2}{3},\frac{2}{3}\right)$   &  $3\left(\lambda_{DN}-\frac{q_{2}}{2}\right)$\\\hline

	NN     &      $1$           	          & $\left(1,0\right)$ 	&   $\lambda_{NN}+\lambda_{PN}-2\lambda_{DD}-2\lambda_{PD}-\lambda_{DN}>0$    \\\hline 

           &      &                                                         &      $\sum = 1$                               \\\hline

    \end{tabular}\label{TableP1SB3}
\end{center}
\end{table*}
In  Table \ref{TableP1SB3}, we present the scheme that achieves the DoF pair corresponding to $P_{1}^{*}$:
\begin{align}
(d_{1},d_{2})&= (1-\lambda_{DD}-\lambda_{PD}-\lambda_{DN}, \lambda_{PP}+2\lambda_{DD}+ 3\lambda_{PD}+2\lambda_{DN}+\lambda_{PN}).
\end{align}
The scheme in Table \ref{TableP1SB3} works for any choice of $(q_{1},q_{2})$ that satisfy the following conditions:
\begin{align}
q_{1}&\leq 3\lambda_{PD}\label{PNB31}\\
q_{2}&\leq 2\lambda_{DN}\label{PNB32}\\
\frac{2q_{1}}{3}+ \frac{q_{2}}{2}&= \lambda_{PN}\label{PNB33}.
\end{align}
The conditions (\ref{PNB31})-(\ref{PNB32}) ensure that the fractions of the constituent schemes are non-negative.
Condition (\ref{PNB33}) ensures that the states (PN, NP) are fully alternated with (PD,DP) and (DN, ND) states and the marginals of the states
are preserved. This is guaranteed by condition (\ref{B3C2}).

The achievable $(d_{1},d_{2})$ are as follows:
\begin{align}
d_{1}&= \lambda_{PP}+\lambda_{NN}+ \lambda_{PD}+\lambda_{DN}+ \lambda_{PN}+ \underbrace{\left(\frac{2q_{1}}{3}+\frac{q_{2}}{2}\right)}_{=\lambda_{PN}}\\
&= \lambda_{PP}+\lambda_{NN}+\lambda_{PD}+\lambda_{DN}+2\lambda_{PN}\\
&= 1-\lambda_{DD}-\lambda_{PD}-\lambda_{DN},
\end{align}
and
\begin{align}
d_{2}&=\lambda_{PP}+ 2\lambda_{PN}+\lambda_{PD}+\lambda_{DN}+ 2\lambda_{DD}+ 2\lambda_{PD}+ \lambda_{DN}-\lambda_{PN}\\
&= \lambda_{PP}+ 2\lambda_{DD}+ 3\lambda_{PD}+ 2\lambda_{DN}+ \lambda_{PN}.
\end{align}

This completes the achievability proof for $\mathcal{D}(\lambda)$ for Case B.

\section{Converse Proofs}\label{Sec:Converse}
\subsection{Proof of $2d_{1}+d_{2}$ upper bound}
We denote the channel output at the receivers as follows:
\begin{align}
Y^{n}&= \left(Y^{n}_{pp}, Y^{n}_{pd}, Y^{n}_{dp}, Y^{n}_{pn}, Y^{n}_{np}, Y^{n}_{dn}, Y^{n}_{nd}, Y^{n}_{dd}, Y^{n}_{nn}\right),\\
Z^{n}&= \left(Z^{n}_{pp}, Z^{n}_{pd}, Z^{n}_{dp}, Z^{n}_{pn}, Z^{n}_{np}, Z^{n}_{dn}, Z^{n}_{nd}, Z^{n}_{dd}, Z^{n}_{nn}\right),
\end{align}
where the subscript $Y^{n}_{ab}$ (respectively $Z^{n}_{ab}$) denotes the portion of the channel output at receiver $1$ (respectively receiver $2$) corresponding to the time instants that transmitter spends in state AB.

We first enhance receiver $2$ by giving it the channel output of receiver $1$, i.e., receiver $2$ now has $(Y^{n},Z^{n})$. For this enhanced physically degraded broadcast channel, it is known from \cite{ElGamalFB} that feedback does not increase the capacity region. Thus, we remove the delayed CSIT assumption from the states PD, DP, DN, ND and DD without effecting the capacity region. 

Now, we introduce a statistically indistinguishable receiver $\widetilde{1}$, which has access to the following channel output:
\begin{align}
\widetilde{Y}^{n}&= \left(Y^{n}_{pp}, Y^{n}_{pd}, \widetilde{Y}^{n}_{dp}, Y^{n}_{pn}, \widetilde{Y}^{n}_{np}, Y^{n}_{dn}, \widetilde{Y}^{n}_{nd}, \widetilde{Y}^{n}_{dd}, \widetilde{Y}^{n}_{nn}\right),
\end{align}
where the channel output to receiver $\widetilde{1}$ is 
\begin{itemize}
\item exactly the same as the channel output at receiver $1$ corresponding to states PP, PD, PN, DN, and
\item identically distributed as the channel output to receiver $1$ in the states DP, NP, ND, DD and NN. 
\end{itemize}

We next note that in this enhanced broadcast channel without feedback, the capacity region  depends only on the marginals.
Therefore, due to this fact and due to the specific construction of the channel output to receiver $\widetilde{1}$, both receivers $1$ and $\widetilde{1}$ can decode the message $W_{1}$. Finally, we also give the output of receiver $\widetilde{1}$ to receiver $2$.

Denote $\Omega= \left(\{  H(i), G(i), \widetilde{H}(i) \}_{i=1}^{n}\right)$ as the global CSIT of the original broadcast channel and the CSIT of the artificial receiver $\widetilde{1}$ for the entire block length $n$.

We thus have the following sequence of inequalities:
\begin{align}
nR_{1}&\leq I(W_{1};Y^{n}|\Omega)+ o(n)\nonumber\\
&= h(Y^{n}|\Omega) - h(Y^{n}|W_{1},\Omega) + o(n)\nonumber\\
&\leq n\log(P) - h(Y^{n}|W_{1},\Omega)+ o(n).\label{Th2-E1}
\end{align}
Similarly, for the artificial receiver $\widetilde{1}$, we have
\begin{align}
nR_{1}&\leq n\log(P) - h(\widetilde{Y}^{n}|W_{1},\Omega)+ o(n).\label{Th2-E2}
\end{align}

Adding (\ref{Th2-E1}) and (\ref{Th2-E2}), we have
\begin{align}
2nR_{1}&\leq 2n\log(P) - h(Y^{n}|W_{1},\Omega)- h(\widetilde{Y}^{n}|W_{1},\Omega)+ o(n)\\
&\leq 2n\log(P) - h(Y^{n}, \widetilde{Y}^{n}|W_{1},\Omega)+ o(n)\label{Th2-E3}.
\end{align}

Now, consider the enhanced receiver $2$:
\begin{align}
nR_{2}&\leq I(W_{2};Z^{n},Y^{n},\widetilde{Y}^{n}|W_{1},\Omega)+ o(n)\nonumber\\
&=h(Z^{n},Y^{n},\widetilde{Y}^{n}|W_{1},\Omega)- \underbrace{h(Z^{n},Y^{n},\widetilde{Y}^{n}|W_{1},W_{2},\Omega)}_{\geq no(\log(P))} +o(n)\nonumber\\
&\leq h(Z^{n},Y^{n},\widetilde{Y}^{n}|W_{1},\Omega)+o(n)-no(\log(P))\nonumber\\
&= h(Y^{n},\widetilde{Y}^{n}|W_{1},\Omega)+ h(Z^{n}|Y^{n},\widetilde{Y}^{n},W_{1},\Omega) + o(n)-no(\log(P))\nonumber\\
&\leq h(Y^{n},\widetilde{Y}^{n}|W_{1},\Omega)+ h(Z^{n}_{pp},Z^{n}_{pd}, Z^{n}_{pn}) + o(n)-no(\log(P))\nonumber\\
&\quad + h(Z^{n}_{dp}, Z^{n}_{np}, Z^{n}_{dn}, Z^{n}_{nd}, Z^{n}_{dd}, Z^{n}_{nn} |Y^{n},\widetilde{Y}^{n},W_{1},\Omega)\nonumber\\
&\leq h(Y^{n},\widetilde{Y}^{n}|W_{1},\Omega)+ n(\lambda_{PP}+\lambda_{PD}+\lambda_{PN}) + o(n)-no(\log(P))\nonumber\\
&\quad + h(Z^{n}_{dp}, Z^{n}_{np}, Z^{n}_{dn}, Z^{n}_{nd}, Z^{n}_{dd}, Z^{n}_{nn} |Y^{n},\widetilde{Y}^{n},W_{1},\Omega)\nonumber\\
&\leq h(Y^{n},\widetilde{Y}^{n}|W_{1},\Omega)+ n(\lambda_{PP}+\lambda_{PD}+\lambda_{PN}) + o(n)-no(\log(P))\nonumber\\
&\quad + \underbrace{h(Z^{n}_{dp}|Y^{n}_{dp},\widetilde{Y}^{n}_{dp},W_{1},\Omega)}_{\leq no(\log(P))} + \underbrace{h(Z^{n}_{np}|Y^{n}_{np},\widetilde{Y}^{n}_{np},W_{1},\Omega)}_{\leq no(\log(P))}\nonumber\\
&\quad + \underbrace{h(Z^{n}_{nd}|Y^{n}_{nd},\widetilde{Y}^{n}_{nd},W_{1},\Omega)}_{\leq no(\log(P))} + \underbrace{h(Z^{n}_{dn}|Y^{n}_{dn},\widetilde{Y}^{n}_{dn},W_{1},\Omega)}_{\leq no(\log(P))}\nonumber\\
&\quad + \underbrace{h(Z^{n}_{dd}|Y^{n}_{dd},\widetilde{Y}^{n}_{dd},W_{1},\Omega)}_{\leq no(\log(P))} + \underbrace{h(Z^{n}_{nn}|Y^{n}_{nn},\widetilde{Y}^{n}_{nn},W_{1},\Omega)}_{\leq no(\log(P))}\label{Th2-Etemp}\\
&\leq h(Y^{n},\widetilde{Y}^{n}|W_{1},\Omega)+ n(\lambda_{PP}+\lambda_{PD}+\lambda_{PN}) + o(n)+no(\log(P))\label{Th2-E4}
\end{align}
where (\ref{Th2-Etemp}) follows from the following facts:
\begin{itemize}
\item $Z^{n}_{dp}$ can be reconstructed within noise distortion from $(Y^{n}_{dp}, \widetilde{Y}^{n}_{dp}, \Omega)$. 
\item $Z^{n}_{np}$ can be reconstructed within noise distortion from $(Y^{n}_{np}, \widetilde{Y}^{n}_{np}, \Omega)$. 
\item $Z^{n}_{dn}$ can be reconstructed within noise distortion from $(Y^{n}_{dn}, \widetilde{Y}^{n}_{dn}, \Omega)$. 
\item $Z^{n}_{nd}$ can be reconstructed within noise distortion from $(Y^{n}_{nd}, \widetilde{Y}^{n}_{nd}, \Omega)$. 
\item $Z^{n}_{dd}$ can be reconstructed within noise distortion from $(Y^{n}_{dd}, \widetilde{Y}^{n}_{dd}, \Omega)$. 
\item $Z^{n}_{nn}$ can be reconstructed within noise distortion from $(Y^{n}_{nn}, \widetilde{Y}^{n}_{nn}, \Omega)$. 
\end{itemize}

Adding (\ref{Th2-E3}) and (\ref{Th2-E4}), and normalizing by $n$, we have
\begin{align}
2R_{1}+R_{2}&\leq \log(P)\left(2+\lambda_{PP}+ \lambda_{PD}+\lambda_{PN}\right)+ o(\log(P)) +\frac{o(n)}{n}.
\end{align}
Dividing by $\log(P)$, and taking the limits $n\rightarrow \infty$ and then $P\rightarrow \infty$, we have the proof for
\begin{align}
2d_{1}+d_{2}&\leq 2+\lambda_{PP}+ \lambda_{PD}+\lambda_{PN}.
\end{align}
The proof for the bound on $d_{1}+2d_{2}$ follows in a similar manner by reversing the roles of receivers $1$ and $2$.

\subsection{Proof of $d_{1}+d_{2}$ upper bound}
We next prove the bound
\begin{align}
d_{1}+d_{2}\leq 1+\lambda_{PP}+ 2\lambda_{PD}+ \lambda_{DD}+ \lambda_{PN}+\lambda_{DN}.
\end{align}
To this end, we denote the channel outputs corresponding to channel states PP, PD, DP, DD collectively as follows:
\begin{align}
Y^{n}_{0}&= \left(Y^{n}_{pp}, Y^{n}_{pd},Y^{n}_{dp},Y^{n}_{dd}\right)\\
Z^{n}_{0}&= \left(Z^{n}_{pp}, Z^{n}_{pd},Z^{n}_{dp},Z^{n}_{dd}\right).
\end{align}
The subscript $0$ denotes the set of states \{PP, PD, DP, DD\}. 
With this notation in place, we can write the channel outputs at the receivers as follows:
\begin{align}
Y^{n}&= \left(Y^{n}_{0}, Y^{n}_{pn},Y^{n}_{np},Y^{n}_{dn},Y^{n}_{nd}, Y^{n}_{nn}\right)\\
Z^{n}&= \left(Z^{n}_{0}, Z^{n}_{pn},Z^{n}_{np},Z^{n}_{dn}, Z^{n}_{nd}, Z^{n}_{nn}\right).
\end{align}

We next enhance the system as follows: whenever the transmitter has delayed CSIT from the receiver, we make it perfect CSIT. In particular, in the enhanced system, 
in the PD, DP, DD states, the transmitter now has perfect CSIT from both receivers, i.e., all four of these states are enhanced to the PP state. Similarly, 
the state DN is enhanced to a PN state, and the state ND is enhanced to a NP state. Note that while we enhance the CSIT availability, the original fractions 
of each of these states are kept the same as they were in the original system.  

Next, for each of the receivers, we introduce another statistically indistinguishable receiver, which cannot reduce the capacity region, and therefore cannot reduce the DoF. Note that now we have 4 receivers, 2 of which, say receivers $1$ and $\tilde{1}$,  wish to decode the message $W_{1}$ and the other two receivers $2$ and $\tilde{2}$, wish to decode the message $W_{2}$. Furthermore, since the capacity depends only on the marginals, without loss of generality we will assume both receivers have the same channels in state $NN$. Starting with this compound setting let us assume full CSIT, which again cannot reduce capacity or DoF. We will prove the DoF outer bound for this compound BC setting.

These channel outputs are summarized as follows:
\begin{center}
    \begin{tabular}{ | c | c | c | c | c | c | c |}
    \hline
    Receiver  &                                   0           &                 PN                             &              NP                          &            DN                                  &   ND                                          &      NN \\ \hline
    $1$   &                         $Y^{n}_{0}$  &       $Y^{n}_{pn}$                      &   $Y^{n}_{np}$                    &             $Y^{n}_{dn}$                &   $Y^{n}_{nd}$                         &       $Y^{n}_{nn}$ \\\hline
    $\widetilde{1}$   &     $Y^{n}_{0}$  &       $Y^{n}_{pn}$                      &   $\widetilde{Y}^{n}_{np}$ &             $Y^{n}_{dn}$               &   $\widetilde{Y}^{n}_{nd}$     &       $Y^{n}_{nn}$ \\\hline
    $2$   &                         $Z^{n}_{0}$  &       $Z^{n}_{pn}$                      &   $Z^{n}_{np}$  		  &             $Z^{n}_{dn}$                &   $Z^{n}_{nd}$                         &       $Y^{n}_{nn}$ \\\hline
    $\widetilde{2}$   &     $Z^{n}_{0}$  &       $\widetilde{Z}^{n}_{pn}$  &   $Z^{n}_{np}$                    &             $\widetilde{Z}^{n}_{dn}$                &   $Y^{n}_{nd}$                         &       $Y^{n}_{nn}$ \\\hline
    \end{tabular}
\end{center}
For the converse, we start with an arbitrary sequence of coding schemes (indexed by $n$) that operate over $n$ channel uses, achieve rates $R_{1}$ and $R_{2}$ for the two receivers, and guarantee that $P_{e}\rightarrow 0$ as $n \rightarrow \infty$.

We now prove the outer bound:
\begin{align}
nR_{1}&\leq I(W_{1}; Y^{n}|\Omega) + o(n) \nonumber\\
&= I(W_{1}; Y^{n}_{0}, Y^{n}_{pn},Y^{n}_{np}, Y^{n}_{dn},Y^{n}_{nd}, Y^{n}_{nn}|\Omega) + o(n)\nonumber\\
&= I(W_{1};Y^{n}_{0}|\Omega, Y^{n}_{pn}, Y^{n}_{np}, Y^{n}_{dn},Y^{n}_{nd},Y^{n}_{nn}) + I(W_{1};Y^{n}_{pn}, Y^{n}_{np}, Y^{n}_{dn},Y^{n}_{nd},Y^{n}_{nn}|\Omega) + o(n)\nonumber\\
&\leq n(\lambda_{PP}+2\lambda_{PD}+\lambda_{DD})\log(P) + I(W_{1};Y^{n}_{pn}, Y^{n}_{np}, Y^{n}_{dn},Y^{n}_{nd},Y^{n}_{nn}|\Omega) + o(n)\nonumber\\
&= n(\lambda_{PP}+2\lambda_{PD}+\lambda_{DD})\log(P)   + o(n)\nonumber\\ 
&\qquad + I(W_{1};Y^{n}_{pn}, Y^{n}_{dn}, Y^{n}_{nn}|\Omega) + I(W_{1};Y^{n}_{np}, Y^{n}_{nd} | Y^{n}_{pn}, Y^{n}_{dn},Y^{n}_{nn}, \Omega) \nonumber\\
&= n(\lambda_{PP}+2\lambda_{PD}+\lambda_{DD})\log(P)   +  I(W_{1},W_{2};Y^{n}_{np}, Y^{n}_{nd} | Y^{n}_{pn}, Y^{n}_{dn},Y^{n}_{nn}, \Omega) + o(n)\nonumber\\ 
&\qquad + I(W_{1};Y^{n}_{pn}, Y^{n}_{dn}, Y^{n}_{nn}|\Omega) - I(W_{2};Y^{n}_{np}, Y^{n}_{nd} | W_{1}, Y^{n}_{pn}, Y^{n}_{dn},Y^{n}_{nn}, \Omega) \nonumber\\
&\leq  n(\lambda_{PP}+2\lambda_{PD}+\lambda_{DD}+ \lambda_{PN}+\lambda_{DN})\log(P)    + o(n)\nonumber\\ 
&\qquad + I(W_{1};Y^{n}_{pn}, Y^{n}_{dn}, Y^{n}_{nn}|\Omega) - I(W_{2};Y^{n}_{np}, Y^{n}_{nd} | W_{1}, Y^{n}_{pn}, Y^{n}_{dn},Y^{n}_{nn}, \Omega) \nonumber\\
&\leq  n(\lambda_{PP}+2\lambda_{PD}+\lambda_{DD}+ \lambda_{PN}+\lambda_{DN})\log(P)    + o(n)\nonumber\\ 
&\quad + I(W_{1};Y^{n}_{pn}, Y^{n}_{dn}, Y^{n}_{nn}|\Omega) - h(Y^{n}_{np}, Y^{n}_{nd} | W_{1}, Y^{n}_{pn}, Y^{n}_{dn},Y^{n}_{nn}, \Omega)\nonumber\\
&\qquad \qquad\hspace{4cm}+ \underbrace{h(Y^{n}_{np}, Y^{n}_{nd} | W_{1}, W_{2}, Y^{n}_{pn}, Y^{n}_{dn},Y^{n}_{nn}, \Omega)}_{\leq no(\log(P))}  \nonumber\\
&\leq  n(\lambda_{PP}+2\lambda_{PD}+\lambda_{DD}+ \lambda_{PN}+\lambda_{DN})\log(P)    + o(n) + no(\log(P))\nonumber\\ 
&\quad + I(W_{1};Y^{n}_{pn}, Y^{n}_{dn}, Y^{n}_{nn}|\Omega) - h(Y^{n}_{np}, Y^{n}_{nd} | W_{1}, Y^{n}_{pn}, Y^{n}_{dn},Y^{n}_{nn}, \Omega)\label{TT-E1}.
\end{align}
Similarly, for receiver $\widetilde{1}$, we have
\begin{align}
nR_{1}&\leq  n(\lambda_{PP}+2\lambda_{PD}+\lambda_{DD}+ \lambda_{PN}+\lambda_{DN})\log(P)    + o(n) + no(\log(P))\nonumber\\ 
&\quad + I(W_{1};Y^{n}_{pn}, Y^{n}_{dn}, Y^{n}_{nn}|\Omega) - h(\widetilde{Y}^{n}_{np}, \widetilde{Y}^{n}_{nd} | W_{1}, Y^{n}_{pn}, Y^{n}_{dn},Y^{n}_{nn}, \Omega)\label{TT-E2}.
\end{align}

\noindent Combining (\ref{TT-E1}) and (\ref{TT-E2}), we obtain
\begin{align}
2nR_{1}&\leq  2n(\lambda_{PP}+2\lambda_{PD}+\lambda_{DD}+ \lambda_{PN}+\lambda_{DN})\log(P)    + o(n) + no(\log(P))\nonumber\\ 
&\quad + 2I(W_{1};Y^{n}_{pn}, Y^{n}_{dn}, Y^{n}_{nn}|\Omega) - h(Y^{n}_{np}, \widetilde{Y}^{n}_{np}, Y^{n}_{nd}, \widetilde{Y}^{n}_{nd} | W_{1}, Y^{n}_{pn}, Y^{n}_{dn},Y^{n}_{nn}, \Omega)\label{TT-E3}.
\end{align}

Now consider the following term appearing in (\ref{TT-E3}):
\begin{align}
&2I(W_{1};Y^{n}_{pn}, Y^{n}_{dn}, Y^{n}_{nn}|\Omega) - h(Y^{n}_{np}, \widetilde{Y}^{n}_{np}, Y^{n}_{nd}, \widetilde{Y}^{n}_{nd} | W_{1}, Y^{n}_{pn}, Y^{n}_{dn},Y^{n}_{nn}, \Omega)\nonumber\\
&= 2I(W_{1};Y^{n}_{pn}, Y^{n}_{dn}, Y^{n}_{nn}|\Omega) \nonumber\\
&\qquad - h(Y^{n}_{np}, \widetilde{Y}^{n}_{np}, Y^{n}_{nd}, \widetilde{Y}^{n}_{nd} , Y^{n}_{pn}, Y^{n}_{dn},Y^{n}_{nn}, | W_{1}, \Omega) + h(Y^{n}_{pn}, Y^{n}_{dn},Y^{n}_{nn}| W_{1},\Omega)\nonumber
\end{align}
\begin{align}
&\leq 2I(W_{1};Y^{n}_{pn}, Y^{n}_{dn}, Y^{n}_{nn}|\Omega) \nonumber\\
&\qquad - h(Y^{n}_{np}, \widetilde{Y}^{n}_{np}, Y^{n}_{nd}, \widetilde{Y}^{n}_{nd} , Y^{n}_{nn}, | W_{1}, \Omega) + h(Y^{n}_{pn}, Y^{n}_{dn},Y^{n}_{nn}| W_{1},\Omega)\nonumber\\ 
&\leq 2I(W_{1};Y^{n}_{pn}, Y^{n}_{dn}, Y^{n}_{nn}|\Omega) + 2no(\log(P))\nonumber\\
&\qquad - h(Y^{n}_{np}, \widetilde{Y}^{n}_{np}, Y^{n}_{nd}, \widetilde{Y}^{n}_{nd} , Y^{n}_{nn}, | W_{1}, \Omega) + h(Y^{n}_{pn}, Y^{n}_{dn},Y^{n}_{nn}| W_{1},\Omega)\nonumber\\ 
&\qquad + \underbrace{h(Y^{n}_{np}, \widetilde{Y}^{n}_{np}, Y^{n}_{nd}, \widetilde{Y}^{n}_{nd} , Y^{n}_{nn} |W_{1},W_{2},\Omega)}_{\leq no(\log(P))}-
\underbrace{h(Y^{n}_{pn}, Y^{n}_{dn}, Y^{n}_{nn}|W_{1},W_{2},\Omega)}_{\geq no(\log(P))}\nonumber\\
&= 2I(W_{1};Y^{n}_{pn}, Y^{n}_{dn}, Y^{n}_{nn}|\Omega) + 2no(\log(P))\nonumber\\
&\qquad - I(W_{2}; Y^{n}_{np}, \widetilde{Y}^{n}_{np}, Y^{n}_{nd}, \widetilde{Y}^{n}_{nd} , Y^{n}_{nn}, | W_{1}, \Omega) + I(W_{2}; Y^{n}_{pn}, Y^{n}_{dn},Y^{n}_{nn}| W_{1},\Omega)\nonumber\\ 
&= I(W_{1};Y^{n}_{pn}, Y^{n}_{dn}, Y^{n}_{nn}|\Omega) + \underbrace{I(W_{1}, W_{2}; Y^{n}_{pn}, Y^{n}_{dn},Y^{n}_{nn}| \Omega)}_{\leq n(\lambda_{PN}+\lambda_{DN}+\lambda_{NN})\log(P)}+ 2no(\log(P))\nonumber\\
&\qquad - I(W_{2}; Y^{n}_{np}, \widetilde{Y}^{n}_{np}, Y^{n}_{nd}, \widetilde{Y}^{n}_{nd} , Y^{n}_{nn}, | W_{1}, \Omega)\nonumber\\ 
&\leq  n(\lambda_{PN}+\lambda_{DN}+\lambda_{NN})\log(P) + 2no(\log(P))\nonumber\\
&\qquad + I(W_{1};Y^{n}_{pn}, Y^{n}_{dn}, Y^{n}_{nn}|\Omega) - I(W_{2}; Y^{n}_{np}, \widetilde{Y}^{n}_{np}, Y^{n}_{nd}, \widetilde{Y}^{n}_{nd} , Y^{n}_{nn}, | W_{1}, \Omega)\nonumber\\ 
&\leq  n(\lambda_{PN}+\lambda_{DN}+\lambda_{NN})\log(P) + 4no(\log(P))\nonumber\\
&\qquad + I(W_{1};Y^{n}_{pn}, Y^{n}_{dn}, Y^{n}_{nn}|\Omega) - I(W_{2}; Z^{n}_{np}, Y^{n}_{np}, \widetilde{Y}^{n}_{np}, Z^{n}_{nd}, Y^{n}_{nd}, \widetilde{Y}^{n}_{nd} , Y^{n}_{nn}, | W_{1}, \Omega)\label{TT-E4}\\ 
&\leq  n(\lambda_{PN}+\lambda_{DN}+\lambda_{NN})\log(P) + 4no(\log(P))\nonumber\\
&\qquad + I(W_{1};Y^{n}_{pn}, Y^{n}_{dn}, Y^{n}_{nn}|\Omega) - I(W_{2}; Z^{n}_{np}, Z^{n}_{nd}, Y^{n}_{nn}, | W_{1}, \Omega)\nonumber\\ 
&\leq  n(\lambda_{PN}+\lambda_{DN}+\lambda_{NN})\log(P) + 4no(\log(P))\nonumber\\
&\qquad + I(W_{1};Y^{n}_{pn}, Y^{n}_{dn}, Y^{n}_{nn},W_{2}|\Omega) - I(W_{2}; Z^{n}_{np}, Z^{n}_{nd}, Y^{n}_{nn}, | W_{1}, \Omega)\nonumber\\ 
&=  n(\lambda_{PN}+\lambda_{DN}+\lambda_{NN})\log(P) + 4no(\log(P))\nonumber\\
&\qquad + I(W_{1};Y^{n}_{pn}, Y^{n}_{dn}, Y^{n}_{nn}|W_{2},\Omega) - I(W_{2}; Z^{n}_{np}, Z^{n}_{nd}, Y^{n}_{nn}, | W_{1}, \Omega),\label{TT-E6}\\
&=  n(1- \lambda_{PP}-2\lambda_{PD}-\lambda_{DD}-\lambda_{PN}-\lambda_{DN})\log(P) + 4no(\log(P))\nonumber\\
&\qquad + I(W_{1};Y^{n}_{pn}, Y^{n}_{dn}, Y^{n}_{nn}|W_{2},\Omega) - I(W_{2}; Z^{n}_{np}, Z^{n}_{nd}, Y^{n}_{nn}, | W_{1}, \Omega),\label{TT-E6b}
\end{align}
where (\ref{TT-E4}) follows from the following facts 
\begin{itemize}
\item $Z^{n}_{np}$ can be reconstructed within noise distortion from $(Y^{n}_{np}, \widetilde{Y}^{n}_{np}, \Omega)$.
\item $Z^{n}_{nd}$ can be reconstructed within noise distortion from $(Y^{n}_{nd}, \widetilde{Y}^{n}_{nd}, \Omega)$,
\end{itemize}
(\ref{TT-E6}) follows from the fact that $W_{1}$, $W_{2}$ and $\Omega$ are all mutually independent random variables, and (\ref{TT-E6b})
follows from the following:
\begin{align}
\lambda_{PP}+2\lambda_{PD}+\lambda_{DD}+2\lambda_{PN}+2\lambda_{DN}+\lambda_{NN}=1.
\end{align}

Substituting (\ref{TT-E6b}) back into (\ref{TT-E3}), we obtain
\begin{align}
2nR_{1}&\leq n(1+ \lambda_{PP}+ 2\lambda_{PD}+\lambda_{DD}+\lambda_{PN}+\lambda_{DN})\log(P)+ o(n)+ 5no(\log(P))\nonumber\\
&\qquad I(W_{1};Y^{n}_{pn}, Y^{n}_{dn}, Y^{n}_{nn}|W_{2},\Omega) - I(W_{2}; Z^{n}_{np}, Z^{n}_{nd}, Y^{n}_{nn}, | W_{1}, \Omega)\label{TT-E7}.
\end{align}
Repeating the same set of arguments for receivers $2$ and $\widetilde{2}$, we obtain
\begin{align}
2nR_{2}&\leq n(1+ \lambda_{PP}+ 2\lambda_{PD}+\lambda_{DD}+\lambda_{PN}+\lambda_{DN})\log(P)+ o(n)+ 5no(\log(P))\nonumber\\
&\qquad I(W_{2};Z^{n}_{np}, Z^{n}_{nd} Y^{n}_{nn}|W_{1},\Omega)  - I(W_{1};  Y^{n}_{pn}, Y^{n}_{dn}, Y^{n}_{nn} |W_{2}, \Omega) \label{TT-E8}.
\end{align}
Adding (\ref{TT-E7}) and (\ref{TT-E8}), we obtain
\begin{align}
2n(R_{1}+R_{2})&\leq 2n(1+ \lambda_{PP}+ 2\lambda_{PD}+\lambda_{DD}+\lambda_{PN}+\lambda_{DN})\log(P) + 2o(n)+ 10no(\log(P)),\nonumber
\end{align}
which upon normalizing by $2n\log(P)$ and taking the limits $n\rightarrow \infty$ and then $P\rightarrow \infty$ yields
\begin{align}
d_{1}+d_{2}&\leq 1+ \lambda_{PP}+ 2\lambda_{PD}+\lambda_{DD}+\lambda_{PN}+\lambda_{DN}.
\end{align}

\section{Conclusions}
A new model of alternating CSIT has been introduced in the context of fading broadcast channels. The DoF region has been characterized for the general alternating CSIT problem. The results highlight the benefits of configurable channel state information; and also reveal the inseparability of these channel states. In practice, the channel availability at the transmitter
can vary dynamically over time and, as our results illustrate in several cases, a complete understanding of the dynamic settings can be easier than the fixed CSIT settings. 
For instance, the individual DoF is not known for the PN (respectively DN) setting. On the contrary, we have obtained the optimal DoF if the states PN and NP (respectively DN and ND) are both present for an equal fraction of the time. The DoF region and claims presented for the alternating CSIT problem are also applicable to the case in which the CSIT at a given time is modeled as an i.i.d. random variable,
where the CSIT state at a given time is $I_{1}I_{2}$ with probability $\lambda_{I_{1}I_{2}}$.  The focus of this paper has been on investigating these dynamic channel conditions and showing their 
benefits for the MISO broadcast channel. We believe that such scenarios are worth investigating for more complicated interference networks, such as the multi-receiver MIMO broadcast, interference and X networks.

\bibliographystyle{unsrt}
\bibliography{refravi}
\end{document}